\documentclass[a4paper,11pt]{article}
\pdfoutput=1

\usepackage{jcappub}
\usepackage[latin1]{inputenc}
\usepackage{amsfonts}  			
\usepackage{amsbsy}  			

\title{The Effective Field Theory of Inflation Models with Sharp Features}
\author{Nicola Bartolo,}
\author{Dario Cannone,}
\author{Sabino Matarrese}
\affiliation{Dipartimento di Fisica e Astronomia ``G. Galilei'', \\ Universit\`a degli Studi di Padova, via Marzolo 8, I-35131 Padova, Italy}
\affiliation{INFN, Sezione di Padova, via Marzolo 8, I-35131 Padova, Italy}
\emailAdd{nicola.bartolo@pd.infn.it}
\emailAdd{dario.cannone@pd.infn.it}
\emailAdd{sabino.matarrese@pd.infn.it}

\abstract{We describe models of single-field inflation with small and sharp step
features in the potential (and sound speed) of the inflaton field, in the context of the Effective Field Theory of Inflation.
This approach allows us to study the effects of features in the power-spectrum and in the bispectrum of curvature perturbations, from a
model-independent point of view, by parametrizing the features directly
with modified ``slow-roll'' parameters. We can obtain a self-consistent power-spectrum, together with enhanced non-Gaussianity, 
which grows with a quantity $\beta$ that parametrizes the sharpness of the step. With this
treatment it is straightforward to generalize and include features in
other coefficients of the effective action of the inflaton field fluctuations. Our conclusion in this case is
that, excluding extrinsic curvature terms, the only interesting effects at the level of the bispectrum could arise from features in
the first slow-roll parameter $\epsilon$ or in the speed of sound $c_s$. Finally, we derive an
upper bound on the parameter $\beta$ from the consistency of the perturbative expansion of the action for inflaton perturbations.  
This constraint can be used for an estimation of the signal-to-noise ratio, to show that the observable which is most sensitive to features is the power-spectrum. 
This conclusion would change if we consider the contemporary presence of a feature and a speed of sound $c_s < 1$, 
as, in such a case, contributions from an oscillating folded configuration can potentially make the bispectrum 
the leading observable for feature models.}

\begin{document}

\newcommand{\dif}{\mathrm{d}}
\newcommand{\me}{\mathrm{e}}
\newcommand{\mi}{\mathrm{i}}
\newcommand{\bs}[1]{\boldsymbol{#1}}

\newcommand{\imgref}[1]{figure \ref{#1}}
\numberwithin{equation}{section}

\maketitle
\flushbottom

\section{Introduction}

One of the goals of modern cosmology is to explain the origin of the observed cosmic structures, which, according to the standard picture, were 
seeded by primordial super-horizon size curvature perturbations, with an almost scale-invariant power-spectrum. 
This picture is strongly supported by the results of the {\it Planck} satellite observations
\cite{planck_overview}, which are in perfect agreement with the simplest models of inflation, in which the potential energy of a single
scalar field drives a period of accelerated expansion in the early Universe. 
However, a wide class of inflationary models allows to fit the same set of data. It is therefore important to explore 
all the possible signatures of inflation, especially in view of the 
current experimental effort. The Effective Field Theory of Inflation (EFTI) \cite{eft:Cheung2007a, eft:Creminelli2006}
(see also~\cite{eft:Weinberg2008}) is an ideal
framework for such an exploration, as it reduces the theory of fluctuations of a single scalar field around an expanding background
to the physics of the Goldstone boson which spontaneously breaks time-diffeomorphism invariance. It would be desirable then to understand if
models that at first sight seem to be not encompassed by the EFTI (such as, for example, resonance models or models with features)
could nevertheless be described within this formalism.

In this paper we focus on the problem of studying single-field models of inflation with sharp features in the inflaton potential in the context of the EFTI. 
The idea of allowing for features in the inflationary potential has a long history 
\cite{feat:Starobinsky1992, feat:Lesgourgues1997, feat:Wang1999, feat:Adams2001, feat:Gong2005},
but it came into vogue several years
later as a possible explanation for the apparent ``glitches'' in the angular power-spectrum of the Cosmic Microwave Background
\cite{feat:Peiris2003, feat:Hunt2004, feat:Covi2006, feat:Joy2007, feat:Joy2008, feat:Mortonson2009, feat:Hazra2010,
feat:Benetti2011, feat:Hazra2012}, as
recently stressed in \cite{feat:Adshead2011}, where it was pointed out that very sharp features could lead to improvements in the
likelihood of WMAP power spectrum.
Beyond the power-spectrum features,
it has been shown that these models generally predict enhanced non-Gaussianity \cite{feat:Wang1999, feat:Chen2006, ng:Chen2008,
feat:Barnaby2010a, feat:Barnaby2010b, feat:Adshead2011}
and can be motivated by some high-energy physics mechanisms \cite{feat:Romano2008, feat:Ashoorioon2006, feat:Ashoorioon2008, feat:Barnaby2009,
feat:Battefeld2010a, feat:Battefeld2010b, feat:Firouzjahi2010, feat:Abolhasani2010, feat:Abolhasani2012, feat:Saito2013, feat:Gao2012}.
All this becomes even more interesting in the light of the recent analyses of {\it Planck} \cite{planck_ps, planck_bs} which, though in complete
agreement with standard featureless slow-roll scenario, have undertaken detailed investigations on feature models. More generally,
features can also be present in the speed of sound \cite{feat:Chen2008, feat:Piazza2008, feat:Nakashima2010, feat:Achucarro2010, feat:Park2012, feat:Achucarro2012a,
feat:Achucarro2012b, feat:Miranda2012, feat:Adshead2013},
giving also in this case characteristic effects on
the power-spectrum, together with possible enhanced non-Gaussianity. It becomes even more attractive then to find a common setup for
this wide phenomenology. As we will show, in the case of very small and very sharp steps in the inflaton potential this is achievable in the context of the EFTI.
This reformulation of feature models will allow us to provide a straightforward generalization to features in the speed of sound or
in every coefficient of higher-dimension operators in the effective Lagrangian. One of the main advantages of our approach is model-independence
and a better understanding of the regime of validity and energy scales involved.

This paper is organized as follows. In section \ref{section1} we briefly review the main concepts behind the EFTI and give the effective
action for single-field models of inflation we will use in the following. 
In section \ref{section2} we show how to describe models with features in the inflaton potential within EFTI and derive, 
using the in-in formalism \cite{ng:Weinberg2005}, the predicted power-spectrum and bispectrum of curvature perturbations. In section
\ref{section3} we generalize the previous approach to include features in other coefficients of the EFTI and show that the most
interesting case is the case of a feature in the speed of sound. Then we compute the power-spectrum and bispectrum for this case, including
the modification to the mode function, which brings an enhanced non-Gaussianity in the folded configurations. In section \ref{section4}
we estimate the signal-to-noise ratio for the modification of the two- and three-point functions.
In appendix \ref{appA} we give a
totally model-independent explanation for the rise of damping effects in the oscillating spectra and bispectra of models with features.
Finally, we compare our results with the existing literature in appendix \ref{appB}.

\section{Effective Field Theory of Single-Field Inflation} \label{section1}

In this section we will briefly review the Effective Field Theory of Inflation (EFTI), describing the main ideas of this approach. The
EFTI we employ in this paper was developed in \cite{eft:Cheung2007a, eft:Creminelli2006} and we refer the reader to those papers for more detailed explanations.

The crucial point in the EFTI is that, instead of writing a Lagrangian for an inflationary model and then studying perturbations around the
FRW expanding background, we are interested in studying the most general effective action of fluctuations directly around a quasi-de Sitter
background, where time-diffeomorphisms are non-linearly realized. The relevant degree of freedom of this effective field theory is then the
Goldstone boson, $\pi$, associated with the spontaneous breakdown of time-diffeomorphism invariance. 
The field $\pi$ will transform non-linearly, $\pi\to\pi-\xi_0(x)$, under time-diffeomorphisms, $t\to t+\xi_0(x)$, and 
parametrizes adiabatic perturbations, i.e. perturbations corresponding to a common local shift in time, $\delta\phi(x)=\phi(t+\pi(x))-\bar{\phi}(t)$. 
Moreover, the field $\pi$ is related (at the linear level) to the comoving curvature perturbations $\zeta$ by
\begin{equation}\label{pizeta_linear}
 \zeta=-H\pi \; .
\end{equation}
The most general effective action for $\pi$ can be first constructed ``geometrically'' using only the metric in a gauge where no scalar
perturbations appear explicitly ($\delta\phi=0$) and then it is translated into the ``$\pi$-language'' by reintroducing the Goldstone boson
through the Stuckelberg trick \cite{eft:Cheung2007a}. At the end, we obtain the Lagrangian
\begin{eqnarray}\label{eft_decoupling}
 S = & \displaystyle \int \dif^4 x \sqrt{-g} & \left[ \frac{1}{2}M_{Pl}^2 R -M_{Pl}^2\dot{H}(t+\pi)\left(\dot{\pi}^2-\frac{{(\partial_i \pi)}^2}{a^2}\right) \right.+\nonumber \\
   & & \left. +\,2M_2^4(t+\pi)\left(\dot{\pi}^2 + \dot{\pi}^3 -\dot{\pi} \frac{{(\partial_i \pi)}^2}{a^2}\right)-\frac{4}{3}M_3^4(t+\pi)\dot{\pi}^3+  .\,.\,. \right]\; ,
\end{eqnarray}
where a dot stands for a derivative w.r.t. cosmic time $t$.

An important simplification occurs when considering the physics of the Goldstone boson at sufficiently high energy, where it decouples
from gravity and can be studied neglecting the mixing with metric fluctuations.
As we are interested in making predictions at energies of order $H$, as long as the decoupling energy scale $E_{mix}$ is smaller than $H$,
we can safely neglect the mixing with gravity. Another important point is the time dependence of the coefficients in eq.~ \eqref{eft_decoupling}. 
The standard approach is to look for solutions where $H$ and
$\dot{H}$ do not vary significantly in time, so that inflation does not cease to be a dynamical attractor, 
although they may have in principle an arbitrary time dependence. From this point of view it is therefore natural to
assume that the same holds for every other coefficient: this makes the Lagrangian approximately invariant under a shift symmetry for $\pi$. 
This kind of reasoning usually brings to neglect every operator that would result from the Taylor expansion of the coefficients,
\begin{equation}\label{taylor}
 f(t+\pi)\simeq f(t)+\dot{f}(t)\pi+\,.\,.\,.\, \; .
\end{equation}
Notice that the shift symmetry on $\pi$ is nothing else that time translation invariance of the inflationary background, that is to say, the
inflationary background is a quasi-de Sitter space. We know that this must be the case at least on time scales of order $H$,
in order to have enough e-foldings to produce the observed spectrum of quasi-scale invariant perturbations. However there
could also be effects that break time translation invariance on time scales much shorter than $H$ and still give an acceptable 
quasi-scale-invariant spectrum. This would be the case, for example, if the slow-roll parameter $\epsilon=-\dot{H}/H^2$, which controls the breaking of shift symmetry,
assumes a temporary modification which however does not violate the necessary condition $\epsilon \ll1$ \cite{res:Behbahani2011}.
We will see in the next section
that this is the starting point for considering inflationary models with step features within the formalism of the EFTI.

\section{Effective Approach for Models with Step Features in the Inflaton Potential} \label{section2}

The common characteristics of models with features are the breaking of the scale-invariance of the power-spectrum and an enhancement
of higher-order correlators, that strongly depends on momenta.
The traditional road followed to deal with models with step features is to specify a form for the inflaton potential $V(\phi)$ and then
study the background evolution of the field, derive expressions for the modified slow-roll parameters and finally study their
effects on the behavior of the correlation functions of curvature perturbations~\nocite{feat:Arroja2011}. 
In this section we want to show how to describe models with step features within the formalism of the Effective Field Theory of Inflation,
studying the effect of features in the Hubble parameter and its derivatives. Let us first restrict to the simplest scenario, 
where all the $M_n(t)$ and $\bar{M}_m(t)$ coefficients of higher-order operators in the effective action~(\ref{eft_decoupling}) are set to zero. 
Consider a potential for the inflaton field of the form \cite{feat:Adshead2011}
\begin{equation}\label{V_feat}
 V(\phi)=V_0(\phi)\left[1+cF\left(\frac{\phi-\phi_f}{d}\right)\right] \; ,
\end{equation}
which describes a step of height $c$ and width $d$ centered at $\phi_f$ with a generic step function $F$. As the field crosses the feature,
a potential energy $\Delta V\simeq c V$ is converted into kinetic energy $\dot{\phi}^2=2 \dot{H}$. As long as the step is small, $c\ll1$,
it does not ruin the inflationary background evolution and its effect can be treated as a perturbation on a standard background.
The idea is then simple: we can describe these models into the EFTI through a time-dependent Hubble parameter $\dot{H}$.
This approach can be easily extended to features in the $M_n(t)$ and $\bar{M}_m(t)$ coefficients of higher-order operators,
as we will show in section~\ref{section3}.

We parametrize the derivative of the Hubble parameter as follows,
\begin{equation}\label{H_feat}
  \dot{H}(t)=\dot{H}_0(t)\left[1+\epsilon_{step}(t)F\left(\frac{t-t_f}{b}\right)\right] \; ,
\end{equation}
which implies that the slow-roll $\epsilon$ will be~\footnote{In order to compare, notice that the parameters
in Eqs. \eqref{epsilon_feat} and \eqref{V_feat} are related by $\epsilon_{step}=3c/\epsilon$, $1/b=H\sqrt{2\epsilon}/d$.}
\begin{equation}\label{epsilon_feat}
  \epsilon=\epsilon_0(t)\left[1+\epsilon_{step}(t)\,F\left(\frac{t-t_f}{b}\right)\right] \; .
\end{equation}
The quantity $\epsilon_{step}$ represents the height of the step, while $t_f$ is its position and $b$ its characteristic width.
The function $F(x)$ goes from $-1$ to $+1$ as its argument passes $x = 0$ with a characteristic width $\Delta x = 1$. We do not give
here any further requirement on the shape of the step and we shall remain as general as possible throughout the paper.
The background parameters $\dot{H}_0(t)$, $\epsilon_0(t)$ and even $\epsilon_{step}$ can in principle have a mild time dependence,
which is controlled by the zeroth-order slow-roll parameters $\epsilon_0$, $\eta_0$, etc. However this variation should be very small in
order not to spoil inflation. Moreover,
we are interested here in the case in which the strongest time dependence comes from the step feature, therefore
we shall take them to be constant in our calculations. It also is clear that, if we want an inflationary background, $\epsilon_{step}$ should
be small, $\left|\epsilon_{step}\right|\ll1$, otherwise we could have a violation of the necessary condition $\epsilon\ll1$.
Provided that, we can expand every quantity in $\epsilon_{step}$ around an unperturbed background, as, e.g.,
\begin{equation}\label{expansion_step}
 \epsilon=\epsilon_0+\epsilon_1+.\,.\,. \; ,
\end{equation}
where dots stand for terms which are higher than first order in $\epsilon_{step}$. Although, as we said, $\epsilon$ is always small, this
could not be the case for higher-order slow-roll parameters, which can temporarily become of order unity or larger.
This happens, for example, for the parameter\footnote{The choice of the second (and higher) order slow-roll parameters is somewhat
arbitrary. Other conventions are possible, for example $\delta=\ddot{H}/2H\dot{H}=-\epsilon-\delta_{ours}$.}:
\begin{equation} \label{delta}
\delta = \frac{1}{2} \frac{\dif \ln \epsilon}{\dif \ln \tau} = -\frac{\dot{\epsilon}}{2\epsilon H}\; ,
\end{equation}
where $\dif \tau=\dif t/a$ is the conformal time. 
We can expand $\delta$ in powers of $\epsilon_{step}$ as
\begin{equation}
\delta=\delta_0 + \delta_1 + \mathcal{O}(\epsilon_{step}^2) \; .
\end{equation}
Notice that this parameter contains a derivative of $\epsilon$ \eqref{epsilon_feat} and hence is proportional to $1/b$, which in principle
can be very large. The major contribution to $\delta_1$ then comes from
\begin{equation}
\delta_1\simeq-\frac{1}{2}\frac{\epsilon_{step}}{H}\dot{F}\left(\frac{t-t_f}{b}\right) \; ,
\end{equation}
This is the situation which we are interested in, as it corresponds to a sharp step feature.
It is useful to rewrite quantities in conformal time. This can be easily done, as we are in a quasi-de Sitter space-time,
\begin{equation}
 \tau\sim-\me^{-Ht} \qquad \Longrightarrow \qquad \frac{t-t_f}{b}=-\beta\ln\frac{\tau}{\tau_f} \; ,
\end{equation}
where $\tau_f$ is the conformal time at which the step occurs and we defined 
\begin{equation}
\beta=\frac{1}{bH}\; .
\end{equation}
Then,
\begin{equation} \label{delta_feat}
 \delta_1=-\frac{1}{2}\epsilon_{step}\beta\,F\,'\left(-\beta\ln\frac{\tau}{\tau_f}\right) \; ,
\end{equation}
where primes denote derivatives with respect to the argument of $F$.

Now we come back to the effective action \eqref{eft_decoupling}, with the Taylor expansion in eq.~\eqref{taylor}. Using eq.~\eqref{H_feat} 
for the time dependence of the Hubble parameter, we obtain an effective theory which can describe models with features
in the inflaton potential. The advantage in using this approach is twofold: first, it becomes easier to identify the regime
of validity of the theory and to assess the relative importance of operators. Second, from this point of view one could easily generalize
feature models to the other couplings in the effective Lagrangian and study all the effects within the same formalism.

\subsection{Power Spectrum}

The first prediction we want to make is the power-spectrum of the curvature perturbations in the case of a sharp step in the inflaton potential 
($\beta\gg1$ i.e. $b\ll1$)..
In order to obtain the equation of motion for the Goldstone boson $\pi$, we need the second-order action, in which the Hubble parameter
is Taylor expanded around $\pi=0$ \cite{res:Behbahani2011}
\begin{equation} \label{S2}
 S_2 = \int \dif^4 x a^3 \left[ -M_{Pl}^2 \dot{H}\left( \dot{\pi}^2-\frac{{(\nabla \pi)}^2}{a^2} \right) + 3 M_{Pl}^2\dot{H}^2 \pi^2\right] \; .
\end{equation}
From the second-order action we derive the equation of motion for $\pi$:
\begin{equation} \label{motion_pi}
 \ddot{\pi}+\left(3H +\frac{\ddot{H}}{\dot{H}}\right)\dot{\pi}-\frac{\nabla^2\pi}{a^2}=
 \ddot{\pi}+H\left(3 -2\delta\right)\dot{\pi}-\frac{\nabla^2\pi}{a^2}=0 \; ,
\end{equation}
where we have neglected a slow-roll suppressed term.
It is easier to discuss the dynamics in conformal time $\dif \tau=\dif t/a$. We can rewrite the action \eqref{S2} in the form
\begin{equation}\label{S2_tau}
 S_2 = \frac{1}{2}\int \dif^3x\dif\tau\, z^2\!\left[\pi'^2-(\nabla\pi)^2-3a^2\dot{H}\pi^2\right] \; ,
\end{equation}
where primes denote differentiation with respect to $\tau$ and
\begin{equation}\label{z}
 z^2=-2a^2M_{Pl}^2\dot{H} \; .
\end{equation}
Making the redefinition $\pi=u/z$, we obtain
\begin{equation}
 S_2 = \frac{1}{2} \int \dif^3 x \dif\tau \left[u'^2-(\nabla u)^2+\left(\frac{z''}{z}+3a^2H^2\epsilon\right)u^2\right] \; .
\end{equation}
Notice that the second derivative of $z$ \eqref{z} contains slow-roll parameters and their derivatives up to the second derivative of
$\epsilon$, which appears through the parameter
\begin{equation}
 \frac{\dot{\delta}}{H}=-\frac{\dif\delta}{\dif\ln\tau} \; .
\end{equation}
It is clear that $\dot{\delta}/H$ will give the largest contribution, being proportional to
$\beta^2$.
To study its effects on curvature perturbations, we look at the equation of motion for $u$ in terms of the variable $x=-k\tau$,
\begin{equation}\label{motion_u}
 \partial_x^2 u -\frac{2}{x^2} u + u=\frac{\dot{\delta}}{Hx^2} \; ,
\end{equation}
where we have neglected some other slow-roll terms, which are much smaller in the case of a small, $\epsilon_{step}\ll1$, and sharp,
$\beta\gg1$, step.\footnote{All other terms are suppressed at least by $1/\beta$, $\epsilon_0$ or $\epsilon_{step}$.}
This equation can be solved using the Green's function technique, treating the right-hand side of eq.~\eqref{motion_u} as a source
function for the left-hand side. The machinery of the
General Slow-Roll (GSR) approximation developed in~\cite{feat:Dvorkin2009,feat:Adshead2011} helps us to accomplish this task and provides us
with a useful formula for the resulting power-spectrum at late times, $\tau\to0$,
\begin{equation}\label{ps_integral}
 \ln\mathcal{P}_\zeta=\ln\mathcal{P}_{\zeta,0}+\frac{2}{3}\int_{-\infty}^{+\infty}\dif\ln\tau\, W(k\tau)\,\frac{\dif \delta}{\dif\ln\tau} \; ,
\end{equation}
where $\mathcal{P}_\zeta=k^3P_\zeta/(2\pi^2)$ and $W(x)$ is the ``window function'':
\begin{equation}
 W(x)=\frac{3\sin(2x)}{2x^3}-\frac{3\cos(2x)}{x^2}-\frac{3\sin(2x)}{2x} \; .
\end{equation}
The zeroth-order power-spectrum is simply
\begin{equation}\label{ps0}
 \mathcal{P}_{\zeta,0}=\frac{H^2}{8\pi^2\epsilon M_{Pl}^2} \; .
\end{equation}
Now, from \eqref{ps_integral}, integrating by parts and using eq.~\eqref{delta_feat} we obtain
\begin{equation}\label{ps_gsr}
 \ln\mathcal{P}_\zeta=\ln\mathcal{P}_{\zeta,0}-\frac{1}{3}\epsilon_{step}\beta\!\int_{-\infty}^{+\infty}\dif\ln\tau\, W'(k\tau)\,F'\left(-\beta\ln(\tau/\tau_f)\right) \; ,
\end{equation}
where
\begin{equation}\label{w'(x)}
 W'(x)=\left(-3+\frac{9}{x^2}\right)\cos(2x)+\left(15-\frac{9}{x^2}\right)\frac{\sin(2x)}{2x}
\end{equation}
is the derivative of $W(x)$ with respect to $\ln x$. \\
Notice that if we take the limit $\beta\to+\infty$, the derivative of the step, $F'(x)$, would become a Dirac delta function. Then the
integration in the previous equation would give a power-spectrum which exhibits constant amplitude oscillations with frequency $2k\tau_f$
up to $k\to+\infty$. As we will see better in the next sections, the limit $\beta\to+\infty$ cannot be taken naively since it is not phyisical, and we must take
into account the finite width of the step. The integral in eq.~\eqref{ps_gsr} can be analitically evaluated when $\beta\gg1$
(see appendix \ref{appA} and refs. \cite{feat:Adshead2011, feat:Stewart2001}), leading to
\begin{equation}\label{ps}
 \ln\mathcal{P}_\zeta=\ln\mathcal{P}_{\zeta,0}-\frac{2}{3}\epsilon_{step} \,W'(k\tau_f)\,\mathcal{D}\left(\frac{k\tau_f}{\beta}\right) \; ,
\end{equation}
where $\mathcal{D}(y)$ is a damping function normalized to one. As shown in appendix \ref{appA}, $\mathcal{D}$ corresponds to the Fourier
transform of the step function $F$ times $(-\mi k)$, irrespective of the particular shape of the step.
We want to stress that this is a general property for models with very sharp steps,
without any further assumptions on the form of the function $F$.
\begin{figure}[t]
\begin{center}
  \includegraphics[scale=0.35]{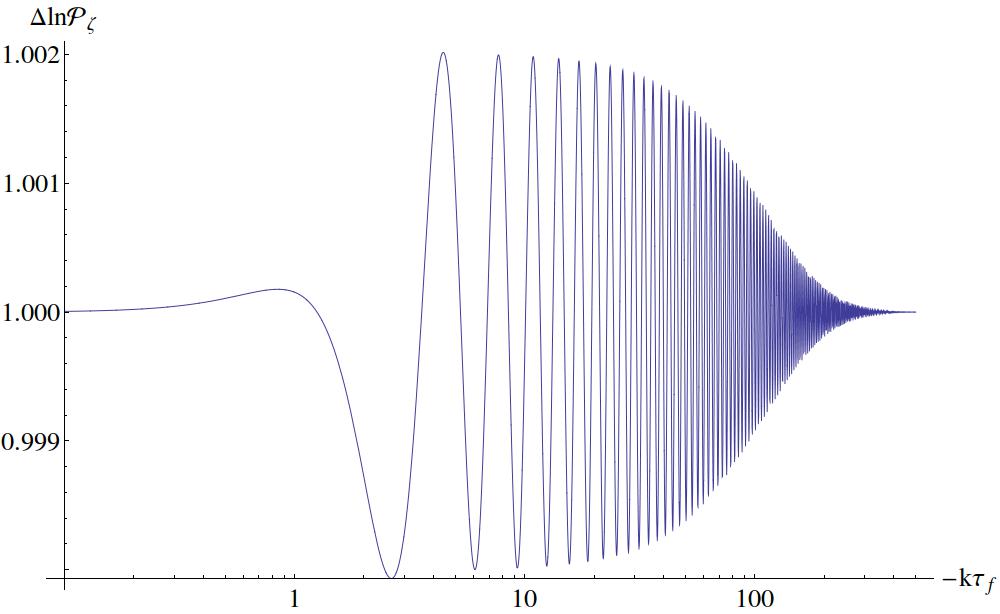}
  \caption{\footnotesize{Non-scale invariant part of the power spectrum \eqref{ps_gsr} for a hyperbolic tangent step \eqref{damp_tanh},
  evaluated for $\epsilon_{step}=0.001$ and $\beta=43\,\pi$ for illustration purposes.}}\label{img_ps}
\end{center}
\end{figure}
Some further comments about eq.~\eqref{ps} are in order. The function $W'(x)$ in \eqref{w'(x)} oscillates between
$-1$ and $+1$ up to $k\to+\infty$ while the function $\mathcal{D}$ acts as a damping envelope. As $x\to0$, $W'(x)\to0$ and no spurious
super-horizon contributions during inflation are generated. Moreover, the damping, decaying exponentially,
``localizes'' the oscillations in an effectively finite range in $k$-space. This was desirable and confirms our intuition that the
feature should not affect modes either much before or much after the step. This is clearly visible from \imgref{img_ps}: the largest
contribution is in the range of frequencies $1\lesssim k\tau_f\lesssim\beta$, which refers to the modes which are inside horizon at the time
of the feature but whose momenta are not greater than the inverse of the time, $b=1/\beta H$, characterizing the sharpness of the step.
It is also clear that, as the parameter $\beta$ becomes
larger, the range in $k$-space in which there are oscillations also becomes larger. In the limit of an infinitely sharp step,
$\beta\to+\infty$, as we already said, the power-spectrum would gain oscillations with constant amplitude up to $k\to+\infty$. Notice finally that the total 
height of the step, namely $2\epsilon_{step}$, does not affect neither the frequency of the oscillations nor the damping and appear
in eq.~\eqref{ps} only as a multiplicative constant in front of the non-scale-invariant part of the spectrum.

\subsection{Bispectrum}

The starting point for computing the bispectrum is the third-order action, which can be derived from eq.~\eqref{eft_decoupling} Taylor
expanding around $\pi=0$:
\begin{equation}\label{S3}
 S_3=\int\dif^4 x a^3 M_{Pl}^2\left[ -\ddot{H}\pi\left(\dot{\pi}^2-\frac{(\nabla \pi)^2}{a^2}\right)-3\dot{H}\ddot{H}\pi^3 \right]\; .
\end{equation}
Notice that we could in principle work in the decoupling regime:
after canonical normalization of the $\pi$ field, $\pi_c=-M_{Pl}\dot{H}^{-1/2}\pi$, we see that we can neglect gravity-mixing interactions
if we work at energies above $E_{mix}\sim \epsilon^{1/2} H$, which is surely below our infrared cutoff $H$, as long as $\epsilon\ll1$.

For the study of non-Gaussianity, we will use the standard in-in formalism (see, e.g.~\cite{ng:Weinberg2005}) and compute the expectation value
\begin{eqnarray} \label{inin}
 \langle \pi_{\bs{k}_1} \pi_{\bs{k}_2} \pi_{\bs{k}_3}\rangle & = & -\mi\,{(2\pi)}^3\delta^3\left(\bs{k}_1+\bs{k}_2+\bs{k}_3\right)
 \int_{-\infty}^0 \dif \tau\,a\, \langle0|\left[\pi_{\bs{k}_1}^{(0)} \pi_{\bs{k}_2}^{(0)} \pi_{\bs{k}_3}^{(0)},
 H_I(\tau)\right]|0\rangle = \nonumber \\
 & = & \mi\,{(2\pi)}^3\delta^3\left(\bs{k}_1+\bs{k}_2+\bs{k}_3\right)\pi_{\bs{k}_1}^{(0)} \pi_{\bs{k}_2}^{(0)} \pi_{\bs{k}_3}^{(0)}
 \int_{-\infty}^0 \frac{\dif \tau}{H^2\tau^2}\,\delta\left(-\beta \ln\tau/\tau_f\right) \times \nonumber \\
 & \times & \pi_{\bs{k}_1}(\tau)^*\bigg[2\pi'_{\bs{k}_2}(\tau)^*\pi'_{\bs{k}_2}(\tau)^*-
 k_1^2\pi_{\bs{k}_2}(\tau)^*\pi_{\bs{k}_3}(\tau)^*\bigg]+\mbox{perm.}+\mbox{c.c.}  \; ,
\end{eqnarray}
where the interaction Hamiltonian $H_I(\tau)$ can be easily read from the third-order action \eqref{S3}. Although also the operator
$\pi^3$ should be present, it can be seen from the action \eqref{S3} that it is proportional to one more factor $\dot{H}$. Therefore its
contribution to the bispectrum will be suppressed by the $\epsilon$ slow-roll parameter.
Notice also that for the computation of this three-point function at leading order we only need the unperturbed mode function
\begin{equation}\label{mode_classic}
 \pi_k^{(0)}(\tau)=\frac{\mi}{M_{Pl}\sqrt{4\epsilon k^3}}\left(1+\mi k \tau\right)\me^{-\mi k \tau} \; .
\end{equation}
As the deviation from the classic solution \eqref{mode_classic} is proportional to $\epsilon_{step}$, its contribution inside
the integral will be suppressed, being at least of order $\mathcal{O}\left(\epsilon_{step}^2\right)$.\footnote{As we will see, this will
be not true if also a speed of sound is taken into account.}
The calculation simplifies using the dimensionless variable $y=z\sqrt{2k} \pi_k$, where $z$ is given by \eqref{z}, which has the form
\begin{equation}\label{mode_y}
 y_0(-k\tau)=\left(1-\frac{\mi}{k\tau}\right)\me^{-\mi k \tau}
\end{equation}
in the unperturbed case. At leading order in the slow-roll parameters and $\epsilon_{step}$ we can evaluate the $\epsilon$ and $H$ factors
inside the integral at horizon crossing and use $\tau\sim-1/aH$.
Then, using the linear relation between $\pi$ and $\zeta$ \eqref{pizeta_linear} we can write
\begin{eqnarray} \label{bs}
 \langle \zeta_{k_1}\zeta_{k_2}\zeta_{k_3}\rangle &=& {(2\pi)}^7 \delta^3\left(\bs{k}_1+\bs{k}_2+\bs{k}_3\right)\frac{\mathcal{P}_{\zeta,0}^2}{4}
 \int_{-\infty}^{0} \frac{\dif \tau}{\tau^2}\,\tau y_0(k_1\tau) \times \nonumber \\
 & & \times \left[2\frac{\dif}{\dif\tau}\big(\tau y_0(k_2\tau)\big)\frac{\dif}{\dif\tau}\big(\tau y_0(k_3\tau)\big)
 -k_1^2 \tau^2 y_0^*(-k_2\tau) y_0^*(-k_3\tau)\right]\delta\left(-\beta\ln\tau/\tau_f\right)+ \nonumber \\
 & & +\mbox{perm}+\mbox{c.c.}
\end{eqnarray}
where we have reconstructed the power-spectrum $\mathcal{P}_{\zeta,0}$ \eqref{ps0} in front of the expression.
This integral is very similar to the one in eq.~\eqref{ps_gsr}
and can be treated in the same way (see appendix \ref{appA}): as we work with very sharp steps, $\beta\gg1$, we can evaluate the polynomials
at $\tau=\tau_f$ so that we are left with the Fourier transform of the step. At the end of the calculation we will obtain an
oscillating function times a damping envelope.
In order to focus on the particular scaling of this type of non-Gaussianity, it is often useful to consider the dimensionless
quantity \cite{feat:Chen2006}:
\begin{equation} \label{G_bs}
 \frac{\mathcal{G}(k_1,k_2,k_3)}{k_1k_2k_3}=\frac{k_1^2k_2^2k_3^2}{{(2\pi)}^4\mathcal{P}_{\zeta,0}^{\,2}}\mathcal{B}(k_1,k_2,k_3) \; ,
\end{equation}
where
\begin{equation}
 \langle \zeta_{k_1}\zeta_{k_2}\zeta_{k_3}\rangle = {(2\pi)}^3 \delta^3(k_1+k_2+k_3)\,\mathcal{B}(k_1,k_2,k_3) \, ,
\end{equation}
or the ``effective'' $\tilde{f}_{\rm NL}$,
\begin{equation}
 \tilde{f}_{\rm NL}(k_1,k_2,k_3)=-\frac{10}{3}\frac{k_1k_2k_3}{k_1^3+k_2^3+k_3^3}\,\frac{\mathcal{G}(k_1,k_2,k_3)}{k_1k_2k_3} \; .
\end{equation}
Since $10k_1k_2k_3/3\sum_ik_i^3$ is roughly of $\mathcal{O}(1)$, the two quantities are of the same order \cite{feat:Chen2006}.
In our case, we find:
\begin{eqnarray} \label{bs3d}
 \frac{\mathcal{G}(k_1,k_2,k_3)}{k_1k_2k_3} &=& \frac{1}{4}\,\epsilon_{step}\,\mathcal{D}\left(\frac{K \tau_f}{2 \beta}\right)
      \Bigg[\Big(\frac{k_1^2+k_2^2+k_3^2}{k_1k_2k_3\,\tau_f}-K\tau_f\Big)K\tau_f\cos(K\tau_f) -\nonumber \\
 & &-\Big(\frac{k_1^2+k_2^2+k_3^2}{k_1k_2k_3\,\tau_f}-\frac{\sum_{i\neq j}k_i^2k_j}{k_1k_2k_3}K\tau_f\Big)\sin(K\tau_f)\Bigg] \; ,
\end{eqnarray}
where $K=k_1+k_2+k_3$.
\begin{figure}[t]
\begin{center}
  \includegraphics[scale=0.35]{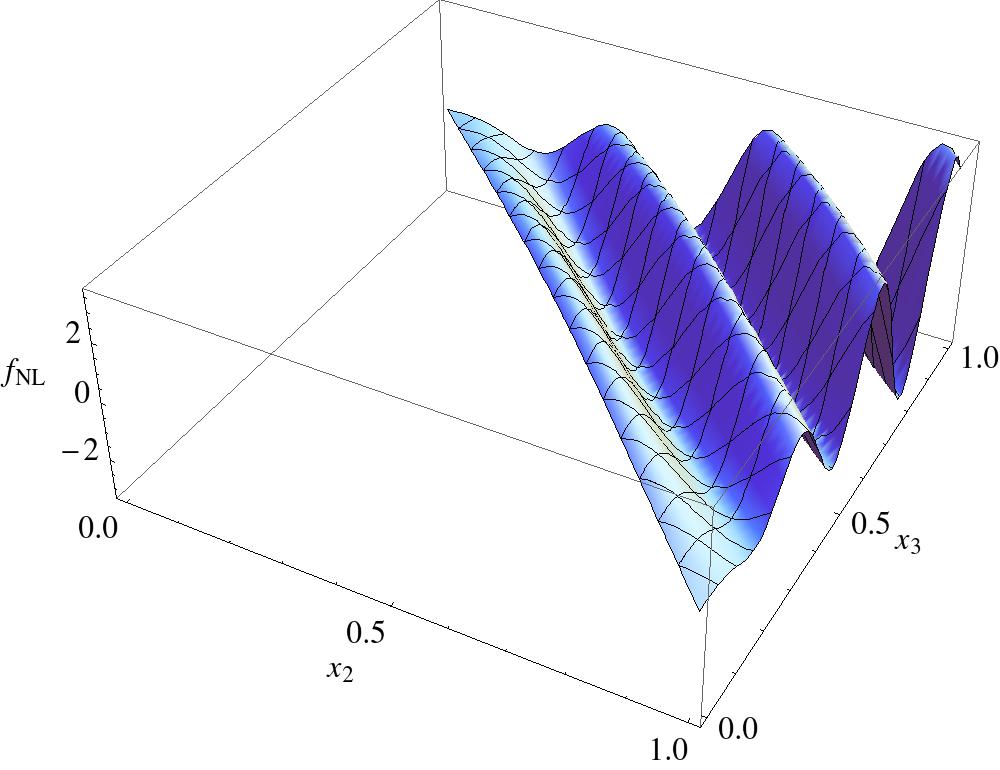}
  \caption{\footnotesize{$\tilde{f}_{\rm NL}(k_1,k_2,k_3)$ for the bispectrum \eqref{bs3d} for a hyperbolic tangent step \eqref{damp_tanh}
  as function of $x_2=k_2/k_1$ and $x_3=k_3/k_1$.
  For illustration purposes we have fixed $k_1=1$ and chosen the values  $\epsilon_{step}=0.001$, $\beta=43\,\pi$ and $\ln(-\tau_f)=3$
  for the parameters.}}
  \label{img_bs3d}
\end{center}
\end{figure}

\subsubsection{Consistency Relation}

It is well known that the bispectrum of curvature perturbations in single-field inflationary models
satisfies a consistency relation which relates its squeezed limit to the slope of the power spectrum \cite{ng:Maldacena2002,
ng:Creminelli2004, eft:Cheung2007b, ng:Creminelli2011} under very general assumptions:~\footnote{See~\cite{ng:Chen2013} and 
refs. therein for a detailed  discussion of the conditions under which one can derive the consistency relation and 
for those cases where one can evade it. See also refs.~\cite{ng:Senatore2012, eft:Creminelli2013}.}
\begin{equation} \label{consistency}
 \lim_{k_L\to0}\mathcal{B}(k_L,k_S,k_S) = -P_\zeta(k_L)P_\zeta(k_S)\left[(n_s-1)
 +\mathcal{O}\left(\frac{k_L^2}{k_S^2}\right)\right] \; .
\end{equation}
In practice, eq.~\eqref{consistency}, which is an expansion in powers of $k_L/k_S$,
tells us that the local physics is unaffected by long-wavelength modes that, being larger than the horizon, cannot be distinguished from
a rescaling of the background. This provides us with a powerful check of our results.
With our notation \eqref{G_bs}, eq.~\eqref{consistency} becomes \cite{feat:Adshead2011}
\begin{equation}
 \lim_{k_L\to0}\frac{\mathcal{G}(k_L,k_S,k_S)}{k_S^3}\simeq-\frac{1}{4}\frac{\dif \ln \mathcal{P}_\zeta}{\dif \ln k}\Bigg|_{k_S} \simeq
 \epsilon_{step} \beta \left(\frac{x}{\beta}\right)\sin(2 x)\,\mathcal{D}\left(\frac{x}{\beta}\right)\, .
\end{equation}
The last equalities comes from the derivative of the power-spectrum \eqref{ps}, neglecting terms of order $\mathcal{O}(1/\beta)$,
i.e. we ignore the variation of the envelope $\mathcal{D}$.
For the squeezed bispectrum, taking the limit $(\bs{k}_2-\bs{k}_3)/2=\bs{k}_S,\,\bs{k}_1=\bs{k}_L\to0$ of eq.~\eqref{bs3d} and focusing on the dominant term,
we find
\begin{equation}\label{bs_squeez}
\lim_{k_L\to0}\frac{\mathcal{G}(k_L,k_S,k_S)}{k_S^3}\simeq\epsilon_{step} \beta \left(\frac{k_S\tau_f}{\beta}\right)
    \sin(2 k_S\tau_f)\,\mathcal{D}\left(\frac{k_S \tau_f}{\beta}\right) \; ,
\end{equation}
which therefore satisfies the consistency relation.
It is important to notice that, for this kind of models, the consistency relation holds only for ``very'' squeezed triangles, that is to
say, here it is not enough to require $k_L/k_S\ll1$. The point is that when we assume that the only effect of the frozen super-horizon
mode on the short wavelength one is a constant background rescaling, we are assuming that there are no interactions between modes when
they are all within the horizon. This is not our case. Here the expansion in $k_L/k_S$ will work only when $k_L$ is sufficiently small
that the mode is already frozen while the short ones are not yet perturbed by the occurrence of the step feature. To derive a bound on $k_L/k_S$, one
can estimates the contribution of the non-Bunch-Davies state to the total energy density and require that it leaves the background
evolution unaltered \cite{eft:Agarwal2012, ng:Flauger2013, ng:Aravind2013}.

\subsubsection{Equilateral Limit and Scaling}

Notice that in the sharp-feature case, $\beta\gg1$, the dominant contribution in eq.~\eqref{bs3d} comes from the terms with
the steepest scaling with $K\tau_f$ and near the equilateral limit (as it can be seen for example in \imgref{img_bs3d}),
where all the momenta are of the same magnitude
\cite{feat:Adshead2011, feat:Adshead2013}. Then we can approximate
\begin{equation}\label{bs_equil}
\frac{\mathcal{G}(k_1,k_2,k_3)}{k_1k_2k_3}\simeq-\epsilon_{step} \beta^2 \left(\frac{K\tau_f}{2\beta}\right)^2
    \cos(K\tau_f)\,\mathcal{D}\left(\frac{K \tau_f}{2 \beta}\right) \; .
\end{equation}
Focusing on the envelope only, we can clearly see that we have a maximum contribution for non-Gaussianity at a scale
$K_{peak}\simeq 2\beta/\tau_f$, which implies: \footnote{Notice that the exact numerical factor in front of eq.~\eqref{fNL_peak}
is model-dependent, as it depends on the normalization of the function $x^2\mathcal{D}(x)$, and hence on the form of the step.}
\begin{equation}\label{fNL_peak}
 f_{\rm NL}\bigg|_{peak}\sim\epsilon_{step}\beta^2\, .
\end{equation}
It is clear however that the bispectrum for these models is strongly scale-dependent both for the oscillating behavior and the envelope.
Then, for arbitrary choice of the parameters, the parameter $f_{\rm NL}(k)$ can change by several orders of magnitude from a scale to
another. This means that an overall amplitude of the oscillations cannot be defined.
We argue then that the ansatz proposed in the papers \cite{feat:Chen2006, ng:Chen2008},
\begin{equation}\label{ansatz_bs}
 f_{\rm NL}^{feat}\sin\left(\frac{K}{k_c}+\phi\right) \; ,
\end{equation}
does not capture the main characteristics of the bispectra of models with very sharp features (as noted also in \cite{feat:Chen2011a}),
if the right damping envelope is not considered.
This approximation loses all the information about the sharpness of the feature,
which actually sets the scale at which modes are most affected. The sharper the feature, the more inside the horizon non-Gaussianity is
produced. Notice also that the ansatz \eqref{ansatz_bs} does not reproduce the correct physical behavior in the limits $K\tau_f\gg1$ and
$K\tau_f\ll1$, since in both cases it does not vanish automatically.
One could solve these problems by hand, multiplying the ansatz \eqref{ansatz_bs} 
by a suitable damping envelope \cite{ng:Chen2010}, at the price of introducing new unknown parameters, where ``suitable''  means that
it must reproduce the correct scaling of eq. \eqref{bs_equil}, with a peak at $K\tau_f\simeq\beta$ and a maximum amplitude given
by eq. \eqref{fNL_peak}. This however will not reproduce correctly the asymmetric
behavior of the envelope \eqref{bs_equil}, which first grows as $K^2$ and then decays exponentially fast.\footnote{Different scaling with
$K$ are possible if the sharp feature is not a step but, e.g., a kink \cite{feat:arroja2012}.} Moreover, in the limit of
an infinitely sharp step, which would correspond to have a very wide damping, we would obtain again oscillations with a constant amplitude, while, as it
has already been noticed in \cite{feat:Adshead2011}, one should obtain a quadratic divergence in momenta space. This behavior is easily
understood: the parameter $\delta$ \eqref{delta_feat} in the limit $\beta\gg1$ is a Dirac-delta function and its only effect in the
integral \eqref{inin} is to replace every $\tau$ with $\tau_f$, without any damping coming into play.
However, as we will see in a moment, this limit can not be taken exactly, if we want to remain in a perturbative regime.

\subsubsection{Regime of Validity}

As we saw, when considering scale dependent bispectra, it is not straightforward to define the equivalent of the constant $f_{\rm NL}$,
characteristic of scale-invariant models. Therefore, if one wants to give an estimate of the strength of non-linear interactions,
it is necessary to specify the energy scale of reference.
This scale will correspond to the energy of the modes that are most affected by the cubic interaction, characteristic
of feature models.

In \cite{feat:Chen2006}, it has been shown that with simple arguments one can estimate a constant $f_{\rm NL}^{feat}$
to be used in the ansatz \eqref{ansatz_bs}, $f_{\rm NL}^{feat}\sim\epsilon_{step}\beta$. This seems to be confirmed by \cite{feat:Adshead2011},
where this value comes from the ratio
\begin{equation}\label{fNL_estim}
\frac{\mathcal{L}_3}{\mathcal{L}_2}\simeq f_{\rm NL}\zeta \; .
\end{equation}
Here we want to show that those estimates implicitly assume that the previous ratio is evaluated at an energy scale of order $H$, that is,
only interactions among modes around horizon crossing are considered.
However, as we have seen, the sharper the feature, the more inside the horizon large interactions
among the modes are effective. The point is that the ratio \eqref{fNL_estim} depends on the energy scale \cite{res:Behbahani2011}.
Using the form of the third-order action \eqref{S3},
\begin{equation}\label{L3/L2}
 \frac{\mathcal{L}_3}{\mathcal{L}_2}\bigg|_{E} = \;\frac{\ddot{H}}{\dot{H}}\pi\bigg|_{E} \; .
\end{equation}
Now we use the fact that $\pi_E$ at an energy scale $E$ is related to $\pi_H$ at Hubble, and hence to $\zeta$ by
\begin{equation}\label{piE}
 \pi_E\sim\frac{E}{H}\pi_H\sim\frac{E}{H^2}\zeta \; .
\end{equation}
Moreover, we know from eq. \eqref{H_feat} the scaling for the time derivatives of the Hubble parameter $H$:
\begin{eqnarray}
 \dot{H} & \sim & \epsilon H^2 \; , \label{scaling1}\\
 H^{(n)} & \sim & \epsilon\, \epsilon_{step} \beta^{n-1} H^{n+1} F\left(\frac{t-t_f}{b}\right) \; . \label{scaling2}
\end{eqnarray}
As the largest interactions happens when the inflaton is crossing the feature, we shall take $t=t_f$. From the previous section, we know
that in the case of a sharp feature, the modes which are most affected are inside horizon, $k\tau_f\sim k/a_f H\sim\beta$, and hence
they have an energy proportional to the inverse of the characteristic time of the feature $1/b=\beta H$. Substituting it into eq.
\eqref{L3/L2} and using eqs. \eqref{piE}, \eqref{scaling1}, \eqref{scaling2}, we find
\begin{equation}
 \frac{\mathcal{L}_3}{\mathcal{L}_2}\bigg|_{E\sim\beta H}\sim\epsilon_{step}\beta^2\, \zeta \; ,
\end{equation}
which is indeed proportional to the $f_{NL}\big|_{peak}$ of eq. \eqref{fNL_peak}.
On the other hand, at energies of order $H$, we have
\begin{equation}
 \frac{\mathcal{L}_3}{\mathcal{L}_2}\bigg|_{E\sim H}\sim\epsilon_{step}\beta\, \zeta \; .
\end{equation}
Once more we see that in the case of sharp features, $\beta\gg1$, the largest interaction comes from modes which are still inside
the horizon at the time of the feature. Therefore, the estimates given in \cite{feat:Chen2006, feat:Adshead2011} implicitly assumes
$E\sim H$.

\begin{figure}[t]
\begin{center}
  \includegraphics[scale=0.35]{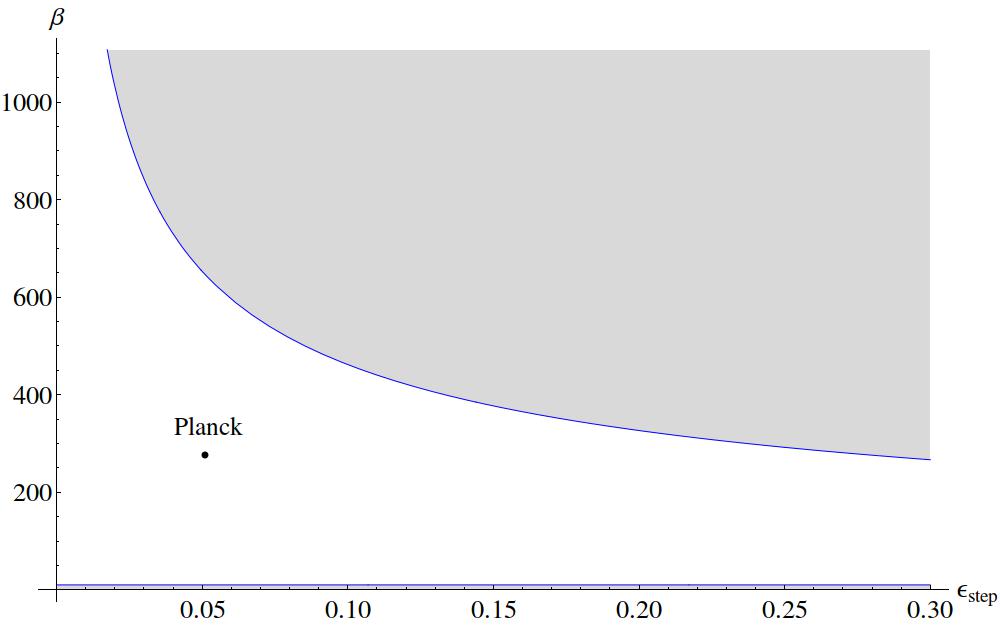}
  \caption{\footnotesize{The shaded region represents the part of the parameter space in which the theory is not under perturbative
  control. In the bottom part of the plot, we have also excluded the region where $\beta\lesssim1$, as it does not correspond to
  a sharp feature. The point represent the best fit value for the {\it Planck} analysis \cite{planck_ps} of power spectra with step features,
  $\{\epsilon_{step},\,\beta\}\simeq\{0.051, 274.4\}$.}} \label{parameterspace}
\end{center}
\end{figure}

Moreover, if we want our theory to be perturbatively safe, we should also require that the ratio $\mathcal{L}_3/\mathcal{L}_2\ll1$.
This ratio indeed gives us an idea of the relative importance of the cubic non-linear interactions with respect to the quadratic ones.
From eq. \eqref{L3/L2} we then derive a bound on $\beta$,
\begin{equation}\label{beta_constraint}
 \beta^2\lesssim \frac{1}{\epsilon_{step}\mathcal{P}_{\zeta,0}^{1/2}} \; .
\end{equation}
This bound (see \imgref{parameterspace}) is stronger than the bound found in \cite{feat:Adshead2011}, as now we are considering the energy scale at which the
largest possible interactions are produced. Violating this bound means that we cannot trust any more our perturbative predictions,
as the contributions from the third-order Lagrangians can be as important as the ones from the second-order Lagrangian
\cite{eft:Baumann2011, eft:Baumann2011c}.\footnote{See also \cite{eft:Avgoustidis2012, eft:Cremonini2010}
for similar bounds derived through adiabaticity-type arguments.}

\section{Generalizations}\label{section3}

Beside features in the inflaton scalar potential, it could be interesting to study possible features, for example,
in the speed of sound \cite{feat:Chen2008, feat:Piazza2008, feat:Nakashima2010, feat:Achucarro2010, feat:Park2012, feat:Achucarro2012a,
feat:Achucarro2012b, feat:Miranda2012, feat:Adshead2013}.
The effective field theory of inflation is the simplest setup for such a study, as it will be a generalization of features in the
$\epsilon$ slow-roll parameter to other coefficients in the effective Lagrangian \eqref{eft_decoupling}.
This can be realized by simply ``switching on'' the coefficients of higher-order
operators that we previously neglected. The coefficients of these new interactions could then
be provided with time dependences in the form of step features, in the same way as we did for the Hubble parameter.

As we saw in the previous sections, if we allow for a time dependence of the $\epsilon$ parameter \eqref{epsilon_feat},
we must also require that its deviation from a constant $\epsilon_0\ll1$ is small in order not to spoil inflation.
This requirement is also necessary to obtain an approximate scale-invariant power-spectrum of curvature perturbations.
In the spirit of the EFTI, a natural explanation is the presence of an
approximate shift symmetry of the Goldstone boson $\pi$, that guarantees that the terms of the Taylor expansion \eqref{taylor} are all
sub-leading with respect to the zeroth-order terms. This conclusion applies to \emph{every} coefficient in the effective action and implies
that every term in the Taylor expansion is completely negligible, including the ones coming from expanding $\dot{H}$. The results of the
previous section, however, tell us that we can still have contributions from the expansion in $\pi$, if the time dependence of the
Hubble parameter assumes a particular form like the one in~\eqref{H_feat}. This effect will still give us an approximate scale-invariant power-spectrum
as long as the shift symmetry still approximately holds, in the sense that it is explicitly broken in a small and controlled way.
Then it is conceivable that also other coefficients in the effective action \eqref{eft_decoupling} could have the same form. This justifies
the generalized study of possible step features in all the parameters appearing in the effective action.
Therefore, we can parametrize the time dependence of the $M_n$ coefficients as
\begin{equation}\label{M_feat}
 M_n(t)=M_n^{(0)}\left[1+m_n F_n\left(\frac{t-t_f}{b_n}\right)\right] \; .
\end{equation}
The meaning of the function $F_n$ and the parameters $m_n$, $t_f$ and $b_n$ are the same of eq. \eqref{epsilon_feat} and we impose
$m_n\ll1$. Now, as we saw for the case of the $\epsilon$ parameter \eqref{epsilon_feat}, interesting new effects arise with operators
that are proportional to the derivative of the step, analogously to eq.~\eqref{delta_feat}, in spite of the step itself, as they are proportional to the
factor $1/b_n$, which can be in principle very large. Looking at the Taylor expansion \eqref{taylor}, we see that the $n$-th derivative
of every coefficient $M_n$, only appears together with $\pi^n$. This means that if a coefficient $M_n$ is present for the first time in
the $m$-th order action, its derivative will appear in the $(m+1)$-order action. As an example, consider the coefficient $M_3$ in the
effective action \eqref{eft_decoupling}, which appears at third order in front of the operator $\dot{\pi}^3$. If it had the time dependence
of eq. \eqref{M_feat}, at third order we would see one more term in the action, which is however proportional to $m_3\ll1$, so that its
contribution would be suppressed with respect to the standard one given by $M_3^{(0)}$. The derivative $\dot{M}_3$, proportional to
$1/b_3\gg1$, which therefore can be large, will appear however with the operator $\pi\dot{\pi}^3$ in the fourth-order action,
that is, its effects must be searched for in the trispectrum.
This leads us to argue that, \emph{at any given order $n>2$ in the effective action, features on a parameter $M_n$ that can be parametrized
by eq. \eqref{M_feat} give
non-negligible effects only if $M_n$ itself has already appeared in the $(n-1)$-th order action}.
Looking at the action of the EFTI and
listing all the terms of the second-order action \cite{eft:Bartolo2010a}, one can see that only the coefficients $\dot{H}$, $M_2$,
$\bar{M}_1$, $\bar{M}_2$, $\bar{M}_3$ are present: this means that only by adding a feature to these coefficients we could hope to see
some feature-related effects at the level of the bispectrum. In practice, we obtain that, neglecting the extrinsic curvature
terms, $\bar{M}_2$, $\bar{M}_3$,
the only interesting effects in the bispectrum can come from features in the Hubble parameter or in the speed of
sound.\footnote{Although also $\bar{M}_1$ is curvature-generated, the corresponding operator is a standard kinetic term and the parameter
can be rewritten as an effective speed of sound for the perturbations \cite{eft:Cheung2007b, eft:Bartolo2010a}. In this paper, we shall
not treat the case of non-vanishing $\bar{M}_2$, $\bar{M}_3$ and leave its study to a future work.}

\subsection{Features in the Speed of Sound}

Focusing for simplicity only on the coefficient $M^4_2(t)$, we can easily see, from
\eqref{eft_decoupling}, that we get a coefficient in front of the time kinetic term which is different from the spatial kinetic one. In
other words, we have a speed of sound
\begin{equation}\label{cs}
 c_s^2(t)=\frac{-M_{Pl}^2 \dot{H}(t)}{-M_{Pl}^2\dot{H}(t) + 2M_2^4(t)} \; .
\end{equation}
Then, it is clear that if we do not neglect the time evolution of the coefficients, we obtain also a time variation of the speed of
sound.\footnote{Notice that even if we do not allow for the time evolution of the coefficient $M_2^4$, we still obtain a time-dependent
speed of sound because of the time-varying Hubble parameter \eqref{H_feat}.}

Here we want to make the example of a step feature in $M^4_2(t)$,
\begin{equation}\label{M2_feat}
 M^4_2(t)=M^4_{2,0}(t)\left[1+\sigma_{step}(t)F\left(\frac{t-t_f}{b_s}\right)\right] \; ,
\end{equation}
where, as in the case of features in the Hubble slow-roll parameters, we could in principle allow for a mild time dependence of the
zeroth-order parameters. Inserting \eqref{M2_feat} into \eqref{cs}, at first order in the parameter $\sigma_{step}$ we find
\begin{equation} \label{cs_feat}
 c_{s,0}^2(t)=c_{s,0}^2\left[1-\sigma_{step}F\left(\frac{t-t_f}{b_s}\right)\right] \; ,
\end{equation}
where $1/c_{s,0}^2=1-2M_{2,0}^2/M_{Pl}^2\dot{H}$. Notice that although the parameter $\sigma_{step}$, $b_s$, $t_f$ and the step function
$F(x)$ have similar physical interpretation as the ones in eq. \eqref{H_feat}, in principle they could be totally different.
For the reasons we have already discussed, we must require that the time variation is small, namely $|\sigma_{step}|\ll1$. This
allows us to expand quantities in the parameter $\sigma_{step}$ as in eq. \eqref{expansion_step}.
Now we can define a ``slow-roll'' parameter,
\begin{equation}\label{sigma}
 \sigma = \frac{\dif\ln c_s}{\dif\ln\tau}=-\frac{\dot{c}_s}{c_s H} \; ,
\end{equation}
which controls the time evolution of the sound speed. If we expand it in powers of $\sigma_{step}$, at first order we have
\begin{equation} \label{sigma_feat}
 \sigma_1 \simeq \frac{1}{2}\sigma_{step}\beta_s\,F\,'\left(-\beta_s\ln\frac{\tau}{\tau_f}\right) \; ,
\end{equation}
where we have switched to conformal time.
The important point here is that this expression is formally equal to the one found for the $\delta$ parameter in eq. \eqref{delta_feat}.

\subsubsection{Power Spectrum}

Following the same steps of the previous sections, in order to study the effects of sharp features in the power-spectrum we start from the second-order
action
\begin{equation} \label{S2_cs}
 S_2 = \int\dif^4x a^3 \left[ -M_{Pl}^2 \dot{H}\left( \frac{\dot{\pi}^2}{c_s^2}-\frac{{(\nabla \pi)}^2}{a^2} \right)
 + 3 \frac{M_{Pl}^2\dot{H}^2}{c_s^2} \pi^2\right] \; .
\end{equation}
The equation of motion for the Goldstone boson $\pi$ reads
\begin{equation} \label{motion_cs}
 \ddot{\pi}+\left(3H +\frac{\ddot{H}}{\dot{H}}-\frac{2\dot{c}_s}{c_s}\right)\dot{\pi}-c_s^2\frac{\nabla^2\pi}{a^2}=
 \ddot{\pi}+H\left(3 -2\delta+2\sigma\right)\dot{\pi}-c_s^2\frac{\nabla^2\pi}{a^2}=0 \; ,
\end{equation}
where we have neglected a slow-roll-suppressed term. Equation \eqref{motion_cs} is formally identical to eq.~\eqref{motion_pi} and
the parameter $\sigma$ defined in \eqref{sigma} enters in the same place as $\delta$ \eqref{delta}. Both the parameters have also the same form
(\eqref{delta_feat}, \eqref{sigma_feat}) at first order in the parameters $\sigma_{step}$ and $\epsilon_{step}$, therefore the main
effect on the power-spectrum will be similar. Setting $\delta=0$ to focus only on the effects of $c_s$, we can follow
the same steps we have followed for the case of a feature in the $\epsilon$ parameter. Switching to conformal time and defining the variable
\begin{equation}\label{z_cs}
z^2=-2a^2M_{Pl}^2\dot{H}/c_s^2 \; , 
\end{equation}
we find an action in the form of eq. \eqref{S2_tau}. Now the second derivative of $z$
contains the parameter
\begin{equation}
 \frac{z''}{z}\supset\frac{\dot{\sigma}}{H}=-\frac{\dif\sigma}{\dif\ln\tau} \; ,
\end{equation}
which gives the dominant contribution in the case of a sharp step, being proportional to $\beta_s^2$. As it can be easily understood,
at this point it is straightforward to write the expression of the power-spectrum at leading order in $\sigma_{step}$:
\begin{equation}\label{ps_cs}
 \mathcal{P}_\zeta=\mathcal{P}_{\zeta,0}\left[1-\frac{2}{3}\sigma_{step}\,\mathcal{D}\left(\frac{k s_f}{\beta_s}\right)
 W'(k s_f)\right] \; ,
\end{equation}
where we have used the variable \cite{feat:Hu2011}
\begin{equation}
 s=\int\frac{c_s \dif t}{a} \; ,
\end{equation}
so that $s_f$ correspond to the time of the feature,
$W'(k s_f)$ is the same oscillating function of eq.~\eqref{w'(x)}, $\mathcal{P}_{\zeta,0}$ is the standard
power-spectrum in the presence of a constant $c_s^2\neq1$:
\begin{equation}
 \mathcal{P}_{\zeta, 0} = \frac{H^2}{8\pi^2 \,\epsilon \,c_{s,0} M_{Pl}^2} \; .
\end{equation}
Again, the damping function $\mathcal{D}$ is nothing else than the Fourier transform of the step itself.
We can see, as already noticed in a previous paper \cite{feat:Miranda2012} for DBI models, that very small and very sharp steps in the
scalar potential or in the speed of sound have strongly degenerate effects on the power-spectrum, as both produce damped oscillations.
If we want to break this degeneracy between the two physically different situations, we have to consider the effects on the bispectrum.

\subsubsection{Bispectrum}

In order to find the effects of the step in the speed of sound, we should consider the action \eqref{eft_decoupling} up to third order in
$\pi$, after Taylor-expanding the coefficients of the various operators. If we focus only on $c_s^2(t+\pi)$, we see that at first order
in $\sigma_{step}$, we get two new operators, namely,
\begin{equation}\label{other_operator}
 -\frac{M_{Pl}^2 \dot{H} }{c_{s,0}^2}\,\sigma_{step} F_s\left(-\beta_s\ln\frac{\tau}{\tau_f}\right) \, \dot{\pi}\left(\dot{\pi}^2-\frac{(\nabla\pi)^2}{a^2}\right)\; ,
\end{equation}
\begin{equation}\label{operator_cs}
 -\frac{2 M_{Pl}^2 \dot{H} }{c_{s,0}^2} H \sigma \,\pi\dot{\pi}^2 \; ,
\end{equation}
where $\sigma$ is given by eq. \eqref{sigma_feat}. Notice that the first of them is just the standard operator present in the EFTI with
speed of sound \cite{eft:Cheung2007a}, times the step function $F_s$ and the parameter $\sigma_{step}$. It is clear then that the
non-Gaussianity produced by this operator will be suppressed by $\sigma_{step}\ll1$ with respect to the standard one, which scales as
$f_{\rm NL}\sim1/c_s^2$. The other operator instead is proportional also to $\beta_s$, which is very large in the case of a sharp step.
To find the corresponding bispectrum, we use the in-in formalism, as in eq. \eqref{inin}. For the leading order result we only need
the zeroth-order mode function,
\begin{equation}\label{mode_cs}
  \pi_k^{(0)}(s)=\frac{\mi}{M_{Pl}\sqrt{4\epsilon c_{s,0}k^3}}\left(1+\mi k s\right)\me^{-\mi k s} \; .
\end{equation}
and the linear relation \eqref{pizeta_linear}. The calculation proceeds along the same path we followed in the case of steps in the Hubble parameter
and the result assumes a similar form
 \begin{eqnarray}\label{bs_cs}
 \frac{\mathcal{G}(k_1,k_2,k_3)}{k_1k_2k_3} &=& \frac{1}{4}\sigma_{step}
 \,\mathcal{D}\left(\frac{(k_1+k_2+k_3)s_f}{2 \beta}\right)\Bigg[-2\sum_{i\neq j}k_ik_js_f^2\cos\big((k_1+k_2+k_3)s_f\big)+ \nonumber \\
 & & +\frac{\sum_{i\neq j}k_i^2k_j}{k_1k_2k_3}\sin\!\big((k_1+k_2+k_3)s_f\big)\Bigg]\; .
\end{eqnarray}
The damping function has the same meaning and properties as the
damping that we have already seen, as it arises from the same kind of integrals (see appendix \ref{appA}). Comparing eqs. \eqref{bs3d} and
\eqref{bs_cs}, we see that, although very similar, the two bispectra can be in principle distinguished both for the different frequency of
the oscillations and for the different combination of momenta $k_1$, $k_2$ and $k_3$.
\begin{figure}[t]
\begin{center}
  \includegraphics[scale=0.4]{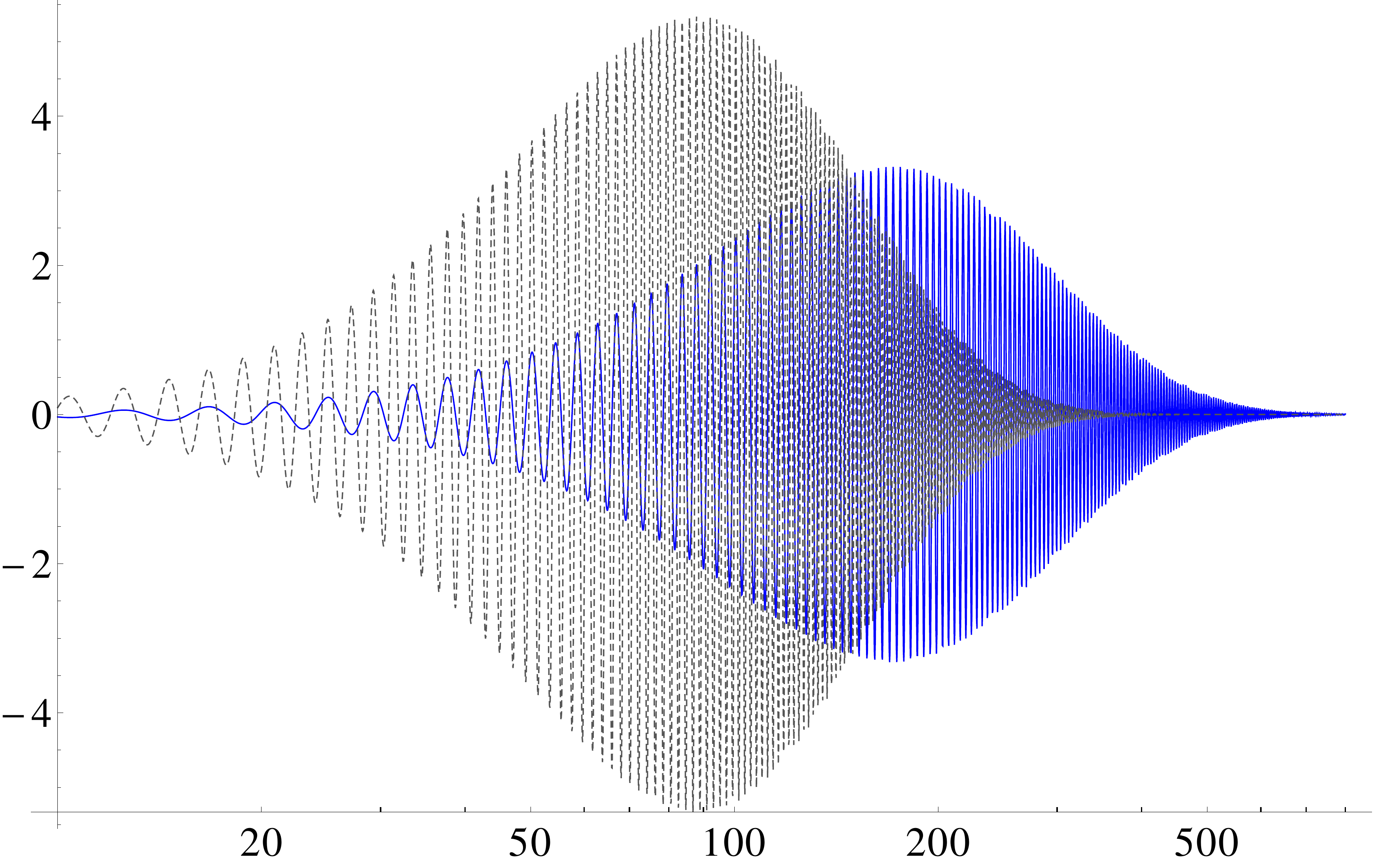}
  \caption{\footnotesize{As an example, we plot the equilateral limit of the bispectra \eqref{bs3d} (blue) and \eqref{bs_cs} (dashed black)
  as functions of $x=-k\tau_f$ in the case of a hyperbolic tangent step \eqref{damp_tanh}.
  The choice of the parameters are $\epsilon_{step}=\sigma_{step}=0.001$, $\beta=\beta_s=43\pi$,
  $c_s=0.5$ for illustration purposes.}}\label{bs_cs_comparison}
\end{center}
\end{figure}
As an example, we show in \imgref{bs_cs_comparison} the behaviour of the two bispectra in the equilateral limit,
$k_1=k_2=k_3=k$, in the case of very small steps in the speed of sound, $\sigma_{step}\ll1$, where the variable $s$ can be approximated
with $s\simeq c_{s,0}\tau$.
Looking at the profiles of the oscillations, we see that both peak at a scale $x_{peak}$, corresponding to the
value $y=1$, where $y$ is the argument of the damping function $\mathcal{D}(y)$. Then, due to the presence of a factor $c_s$, we have
that the first profile peak at $x_{peak}$, while the second at $x_{peak}^{c_s}=x_{peak}/c_s$, which is bigger than
$x_{peak}$ as long as $c_s<1$. Therefore we expect that the two physically different cases are well distinguishable as we move from $c_s=1$ to smaller
values. This conclusion is also reinforced by the non-negligible presence of the characteristic operators of the
small speed of sound scenario, namely the operators proportional to $(1-1/c_s^2)$ in the effective action of eq.
\eqref{eft_decoupling}.\footnote{The bispectrum of the operators $\dot{\pi}^3$ and $\dot{\pi}(\nabla\pi)^2$ gives just the well-known
results for an inflaton with a non-standard kinetic term (see for example \cite{ng:Chen2006}), since correcting the coefficient $(1-1/c_s^2)$ with its step-like evolution
will give only the $\sigma_{step}$-suppressed operator \eqref{other_operator}, which is negligible at first order.} Moreover, as we will
see in the next subsection, we should also consider now the correction to the mode functions that we previously neglected.
Notice finally that, as in the case of bispectra for features in $\epsilon$, we can recast eq. \eqref{bs_cs} in a form which involves
only an integral in $K=k_1+k_2+k_3$ times some combinations of the momenta. This is shown in appendix \ref{appB}.

Also in this case we can derive a strong bound on the parameter $\beta_s$ from the perturbative validity of the theory. In this case, as
we have seen, the modes which are most affected by the interaction have energies of order $E\sim\beta_s H/c_s$, since the bispectrum
peaks around $c_{s,0}k\tau_f\sim\beta_s$. Using the same technique of eq. \eqref{L3/L2}, we find
\begin{equation}\label{beta_cs_constraint}
 \beta_s^2\lesssim\frac{c_{s,0}^2}{\sigma_{step}\mathcal{P}_{\zeta,0}^{1/2}} \; ,
\end{equation}
which is stronger as we move away from $c_{s,0}=1$.

\subsection{Accounting for a non-Bunch-Davies wave function: Folded Shape}

Another interesting source of features in the bispectrum comes from the correction to the classical Bunch-Davies mode. So far, we have
considered only the standard Bunch-Davies mode \eqref{mode_classic} in the computation \eqref{inin}, as deviations enter with a factor
$\epsilon_{step}$. Thus, as the dominant cubic operators are already proportional to $\epsilon_{step}$, the contribution would be suppressed.
However, when considering for example speeds of sound different from one, we have also cubic operators which are zeroth-order in
$\epsilon_{step}$. As they are enhanced by $c_s^{-2}$ in the case of small speed of sound, the effects of a non-Bunch-Davies wave function
due to the presence of features can become relevant, as noted in \cite{feat:Adshead2013}.
This holds even more in general for every operator in the effective action which
is zeroth-order in the parameter that controls the deviation from Bunch-Davies and happens both if we
have features in the slow-roll parameters or in the speed of sound. In the presence of large interactions, these contributions to the
non-Gaussianity can have a comparable size with the previously considered case. The main characteristic of this kind of non-Gaussianity
is its enhancement in the folded triangle limit due to the presence of the negative-frequency mode.

In order to see how this mechanism works, we will compute the bispectrum arising from the operator $\dot{\pi}^3$ in the effective action
\eqref{eft_decoupling} in the case of a sharp step in the slow-roll parameter $\epsilon$ and a constant speed of sound $c_s^2<1$.
As we saw, the second order action gives us
the equation of motion \eqref{motion_cs}, where the non-negligible effect of the parameter $\delta$ results in a modification of the
standard mode function \eqref{mode_cs}. This is indeed the physical interpretation of the oscillations in the power-spectrum:
the feature excites a non-Bunch-Davies component with negative frequency \cite{feat:Chen2006, feat:Chen2008, ng:Chen2010}.
The contribution of this modification to the wave function in the calculation of the three-points functions at first order in
$\epsilon_{step}$ is obtained substituting one of the three positive-frequency mode which enter the integration in the in-in formalism
with a negative-frequency one, $u_{-}\sim\me^{-\mi x}$ (and summing over the different possible choices of this negative-frequency mode):
\begin{eqnarray}\label{inin_folded}
 \langle \pi_{k_1}\pi_{k_2}\pi_{k_3}\rangle & = & {(2\pi)}^3 \delta^3\left(\bs{k}_1+\bs{k}_2+\bs{k}_3\right)\,
 \left[\prod_{i=1}^3\frac{-\mi}{4M_{Pl}^2\epsilon c_s k_i^2}\right]\frac{1}{H} \times \nonumber \\
 & & \times \int_{-\infty}^{0}  \frac{\dif\tau}{\tau}\left[\frac{\dif}{\dif\tau}\left(\tau y^{NB}(-k_1\tau)^*\right)\frac{\dif}{\dif\tau}
     \left(\tau y(-k_2\tau)^*\right)\frac{\dif}{\dif\tau}\left(\tau y(-k_3\tau)^*\right)\right]+\nonumber \\
 & & +\mbox{c.c}+\mbox{perm.}+\mbox{other choices of } y^{NB} \; ,
\end{eqnarray}
where we used the dimensionless variable $y=\sqrt{2k}u_k=z\sqrt{2k}\, \pi_k$ and $z$ is given by eq. \eqref{z_cs}.
The superscript ``NB'' refers to the negative-frequency contribution. At first order in $\epsilon_{step}$ it can be computed solving
the equation of motion \eqref{motion_cs} through the Green's Function technique \cite{feat:Dvorkin2009, feat:Hu2011, feat:Adshead2013}:
\begin{equation}
 y_{NB}(-k \tau)=-\mi y_0^*(-k \tau)\int_{-\infty}^{\tau}\frac{\dif \tau'}{\tau'}\left(1-\frac{\mi}{c_s k \tau'}\right)^2\,
 \frac{\me^{-2\mi c_s k \tau'}}{2c_s k \tau'}\frac{\dif\delta}{\dif\ln\tau'} \; ,
\end{equation}
where $y_0$ is given by eq. \eqref{mode_y}. After some lengthy algebra and an integration by parts, we are left in eq.~(\ref{inin_folded}) with the evaluation
of integrals similar to those of the previous sections, where an oscillating exponential multiplies a polynomial in $k \tau$.
Using again the same technique of appendix \ref{appA}, we end up again with a bispectrum in the form of a oscillating function times a
damping envelope. Instead of the full result, it is easier to focus only on the dominating factor, proportional to $\tau_f$,
\begin{eqnarray}\label{flat3d}
 \frac{\mathcal{G}(k_1,k_2,k_3)}{k_1k_2k_3} & = & \epsilon_{step}\left(1-\frac{1}{c_{s}^2}\right)\Bigg[\frac{3}{k_1k_2k_3}
 \frac{\sum_{i\neq j}k_i^4k_j^2-2\sum_{i\neq j}k_i^3k_j^3-3k_1^2k_2^2k_3^2}{(k_1-k_2-k_3)(k_1-k_2+k_3)(k_1+k_2-k_3)}\Bigg]\times\nonumber \\
 & & \times \mathcal{D}\left(\frac{K c_s \tau_f}{2 \beta}\right) K c_s \tau_f \cos\big(K c_s\tau_f\big) \; .
\end{eqnarray}
\begin{figure}[t]
\begin{center}
  \includegraphics[scale=0.35]{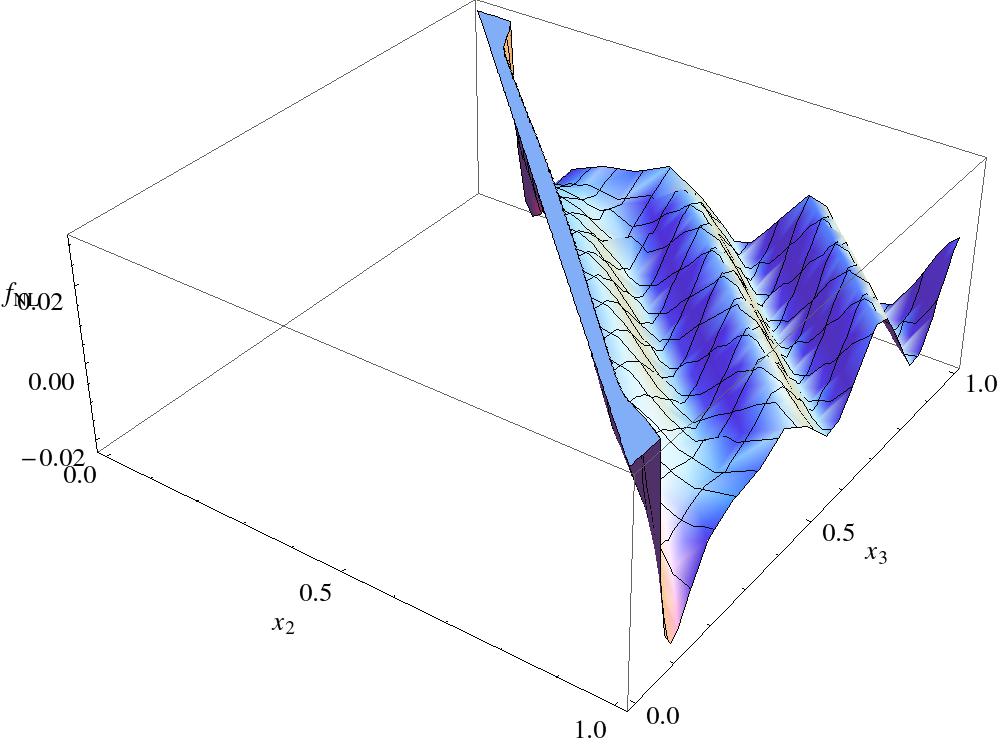}
  \caption{\footnotesize{$\tilde{f}_{\rm NL}$ for the bispectrum \eqref{inin_folded} for a hyperbolic tangent step \eqref{damp_tanh}
  as function of $x_2=k_2/k_1$ and $x_3=k_3/k_1$.
  For illustration purposes we have fixed $k_1=1$ and chosen the values  $c_s=0.4$, $\epsilon_{step}=0.001$, $\beta=43\,\pi$ and $\ln(-\tau_f)=4$
  for the parameters.}}
  \label{img_flat3d}
\end{center}
\end{figure}
\noindent As it can be seen from \imgref{img_flat3d}, the bispectrum \eqref{flat3d} peaks in the folded limit $k_1\to k_2+k_3$,
as it should, given the negative-frequency correction to the mode functions, and has superimposed oscillations similar to those found
for resonant models \cite{res:Behbahani2011, res:Chen2010}. To see the running of this bispectrum in the folded limit,
one has to go back to eq. \eqref{inin_folded} and take $k_1=k$, $k_2\to k/2$, $k_3\to k/2$. Focusing on the dominant contribution,
that is with the steepest scaling with $x=-k\tau$, we find
\begin{equation} \label{bs_cs_flat}
 \frac{\mathcal{G}(k_1,k_2,k_3)}{k_1k_2k_3}\Bigg|_{\mbox{\footnotesize{folded}}}=-\frac{1}{2}\epsilon_{step}\left(1-\frac{1}{c_{s}^2}\right)\,
 \mathcal{D}\left(\frac{c_s x}{\beta_s}\right)c_{s}^2 x^2\sin(2c_{s}x) \; .
\end{equation}
Here, the maximum is at a scale $c_sk\tau_f\sim\beta$ and reach the value
\begin{equation}\label{fNL_peak_flat}
 f_{\rm NL}^{folded}\Bigg|_{peak}\sim\epsilon_{step}\beta^2\left(1-\frac{1}{c_{s}^2}\right) \; .
\end{equation}
This differs significantly from the maximum non-Gaussianity estimated in eq. \eqref{fNL_peak}, where the speed of sound was exactly $c_s=1$.
Now, in the folded limit and at the scale $k$ where the bispectrum peaks, we receive a further enhancement proportional to $1/c_s^2$.
As it has been noted for resonant models \cite{res:Chen2010}, also for feature models the folded bispectrum can be less, equally or more
important than the feature bispectrum depending on the values of the parameters.

\section{Signal to Noise Ratio}\label{section4}

As a last step, let us make some considerations on the observability of features either in the power spectrum or in the bispectrum.
As we saw, one of the most interesting characteristic of models with features is the fact that their effects in both these observables
depend on the same set of parameters, which gives, in principle, the possibility to constrain them at the same time with two independent
analyses. It could be interesting then to ask in which observable we should expect to see a stronger signal.
To answer this question, let us estimate the signal-to-noise ratio as function of the parameters of the model.
Following \cite{feat:Adshead2011, ng:Hu2000}, we write:
\begin{eqnarray}
 \left(\frac{S}{N}(\delta\langle\zeta^2\rangle)\right)^2 & \simeq & 2\pi\int\frac{\dif^2l}{(2\pi)^2}\left(\frac{\delta C_l}{C_l}\right)^2 \; , \\
 \left(\frac{S}{N}(\langle\zeta^3\rangle)\right)^2 & \simeq & 4\pi\int\frac{\dif^2l_1}{(2\pi)^2}\int\frac{\dif^2l_2}{(2\pi)^2}
  \frac{B_{(l_1,l_2,l_3)}^2}{6C_{l_1}C_{l_2}C_{l_3}} \; ,
\end{eqnarray}
where
\begin{eqnarray}
 C_l & = & \frac{1}{5^2D^2}\int\frac{\dif k_{1||}}{2\pi}\mathcal{P}_\zeta(\bs{k_1}) \; , \\
 B_{(l_1,l_2,l_3)} & = & -\frac{2}{5^3D^4}\int\frac{\dif k_{1||}}{2\pi}\int\frac{\dif k_{2||}}{2\pi}\mathcal{B}(\bs{k}_1,\bs{k}_2,\bs{k}_3)
\end{eqnarray}
and $\bs{k}_{1,2}=(\bs{l}_{1,2}/D,k_{1,2||})$, $\bs{k}_3=-\bs{k}_1-\bs{k}_2$, $D$ is the distance to recombination and $||$ is the
direction along the line of sight.
For the explicit calculation, let us focus for simplicity on the case of a feature in the $\epsilon$ slow-roll parameter in the form of
hyperbolic tangent (see appendix \ref{appA}). From eqs. \eqref{ps}, \eqref{bs_equil} we can roughly approximate the maximal signal to
noise accessible to CMB experiments in terms of $\epsilon_{step}$, $\beta$, and $\tau_f$ \cite{feat:Adshead2011}:
\begin{eqnarray}
 \left(\frac{S}{N}(\delta\langle\zeta^2\rangle)\right)^2 & \simeq & 2\pi\epsilon_{step}^2\left(\frac{D}{|\tau_f|}\right)l_{max} \; ,\label{ratio1}\\
 \left(\frac{S}{N}(\langle\zeta^3\rangle)\right)^2 & \simeq & 480\, \epsilon_{step}^2 \left(\frac{\tau_f}{\mbox{Gpc}}\right)^2
  \left(\frac{l_{max}}{2000}\right)^4 \; . \label{ratio2}
\end{eqnarray}
Here $l_{max}$ is the maximum multipole beyond which the signal-to-noise ratio saturates. This is set either by the resolution of the
experiment or by the damped behaviour of our predicted observables. In fact we have seen that the amplitude of the spectrum and bispectrum
is exponentially damped away for high $k$, which means that there is an effective maximum multipole beyond which the signal is
strongly suppressed,
\begin{equation}
 l_d\simeq\frac{2D\beta}{\pi|\tau_f|} \; .
\end{equation}
Therefore we chose $l_{max}$ to be the smallest values between the damping scale $l_d$ and the resolution limit $l_{res}$ that we fix at $l_{res}\simeq2000$.
Now we make the ratio between eqs. \eqref{ratio1}, \eqref{ratio2}  to compare the signals from the modifications of the two-point function
and the three-point function. In the case $l_d<2000$ we find:
\begin{equation}
 \frac{\displaystyle\left(\frac{S}{N}(\langle\zeta^3\rangle)\right)}{\displaystyle\left(\frac{S}{N}(\delta\langle\zeta^2\rangle)\right)}\simeq\,10^{-5}\,\beta^{3/2} \; .
\end{equation}
Using the bound \eqref{beta_constraint} for consistency of the perturbation expansion, we obtain the interesting result that:
\begin{equation}\label{limit1}
 \frac{\displaystyle\left(\frac{S}{N}(\langle\zeta^3\rangle)\right)}{\displaystyle\left(\frac{S}{N}(\delta\langle\zeta^2\rangle)\right)}\lesssim1 \qquad
  \mbox{unless} \quad \epsilon_{step}\lesssim10^{-3} \; .
\end{equation}
On the other hand, if we take $l_{max}=2000$, we obtain:
\begin{equation}\label{limit2}
\frac{\displaystyle\left(\frac{S}{N}(\langle\zeta^3\rangle)\right)}{\displaystyle\left(\frac{S}{N}(\delta\langle\zeta^2\rangle)\right)}
  \simeq 10^{-3} \left(\frac{|\tau_f|}{\mbox{Gpc}}\right)^3 \lesssim1 \qquad \mbox{unless} \quad |\tau_f|\gtrsim10 \,\mbox{Gpc} \; .
\end{equation}
However, as we chose $l_{max}=2000$, we have
\begin{equation}
 l_d\simeq\frac{2D\beta}{\pi|\tau_f|}>2000 \; .
\end{equation}
Then, violating the inequality \eqref{limit2} requires at the same time
\begin{equation}
 10\,\mbox{Gpc}\lesssim|\tau_f|\lesssim10^{-3}\frac{D\beta}{\pi}
\end{equation}
which requires in turn that
\begin{equation}
 \beta\gtrsim10^4 \pi\left(\frac{D}{\mbox{Gpc}}\right)^{-1} \; .
\end{equation}
One more time, looking at the bound of eq. \eqref{beta_constraint}, we obtain that this can happen only for very small values of the
$\epsilon_{step}$ parameter. However, as $\epsilon_{step}$ becomes smaller, we also expect that the signal-to-noise ratio of the bispectrum
itself will become smaller. This can be seen from eq. \eqref{ratio2}:
\begin{equation}
\left(\frac{S}{N}(\langle\zeta^3\rangle)\right)^2\lesssim480 \,\epsilon_{step}^2 \left(\frac{\tau_f}{\mbox{Gpc}}\right)^2
  \lesssim 480\,\epsilon_{step}^2 \left(\frac{D}{\mbox{Gpc}}\right)^2 \; . 
\end{equation}
The last inequality comes from the cutoff $|\tau_f|\lesssim D$ imposed by the projection onto the spherical sky \cite{feat:Adshead2011}.
It is clear then that we would not have a signal-to-noise ratio larger than one if we have $\epsilon_{step}\lesssim10^{-3}$.

This means that the most sensitive test for these models is the feature part of the power spectrum, unless the height of the step is
extremely small, so that one can increase the value of the sharpness of the feature without violating the bound \eqref{beta_constraint}.
However, in this case, it would be too hard to detect any feature effect, as the signal-to-noise would be very small.
This conclusion remains valid if we generalize to features in the speed of sound, where we have an even stronger bound
\eqref{beta_cs_constraint} as we move away from $c_s=1$. The only case that could in principle escape this conclusion would be the case of
folded non-Gaussianity.
Those configurations can potentially make the three-point function the leading observable for feature models since,
for particular choices of the parameters, the folded bispectrum can become dominant and enhance the signal-to-noise ratio. This can be
understood also focusing on the parametric scaling of $f_{NL}\big|_{peak}$ of eq. \eqref{fNL_peak_flat}, which is proportional to
$\beta^2$ but also to $1/c_s^2$, receiving then a further enhancement.

\section{Conclusions}

In this work we have studied models with features, using the approach of the Effective Field Theory of Inflation. We derived predictions
for the power-spectrum and bispectrum for models where a feature is present in the slow-roll parameter $\epsilon$ or in the speed of
sound $c_s$. The starting consideration is that on very small time-scales the background evolution could be very different from de Sitter,
as long as the deviation is small enough to preserve inflation and soon comes back to the attractor solution. As a step feature in the 
potential of the inflaton translates into a similar feature in the slow-roll parameter $\epsilon=- \dot{H}/H^2$, we can describe these models in the EFTI
giving a specific form to the time-variation of the Hubble parameter and its derivatives. This is valid in the case of a very small and
very sharp step. Here, ``small'' means that the total deviation of the slow-roll parameter $\epsilon(t)$ from the constant $\epsilon_0\ll1$ must
be controlled by a parameter $\epsilon_{step}$, which is indeed related to the height of the step, while ``sharp'' means that the characteristic
time-scale of the variation, $b$, should be much smaller than the characteristic time, $H^{-1}$, of inflationary evolution.
This means that the parameter characterizing the sharpness of the step feature is $\beta=1/(bH)\gg1$. 
Under these assumptions, it is possibile to analytically compute the effects of features in the power-spectrum
and bispectrum. These effects are larger for modes still inside horizon at the
time of the feature. We also showed that the expected amplitude of non-Gaussianity at the scale where it receives the largest enhancement
is different from estimates in the previous literature. 

Our technique also allows for a straightforward generalization to include possible features in other coefficients of the EFTI Lagrangian.
Very interestingly, we found that in this case at the level of the three-point function, excluding curvature terms, the only interesting scenario is the one
of a feature in the speed of sound. Then, we have computed the power-spectrum and bispectrum for this scenario, showing that the degeneracy
between the cases of steps in the potential and the speed of sound, which is present in the power-spectrum, can in principle be removed
at the bispectrum level.

In models with resonant non-Gaussianity there is also the possibility of having an enhanced bispectrum in the folded limit. This happens
when considering the non-Bunch-Davies mode functions in the calculation of the three-point correlators. Also in the case of models
with features it is possible to encounter such a situation. The temporary deviation from slow-roll evolution due to the step,
induces a modification of the mode function which is naturally proportional to the height of the step feature $\epsilon_{step}$.
Therefore, any contribution of this modification in operators that are already proportional to $\epsilon_{step}$ is suppressed at leading
order. In the presence of a speed of sound $c_s < 1$, however, there are operators which are zeroth-order in $\epsilon_{step}$. The
corrected mode function there can give contributions to non-Gaussianity which can be less, equally or more important depending on
the parameters of the model and exhibits an oscillating folded shape, analogously to the resonant case.

Finally, the study of the energy-scale of the modes most affected by non-linear interactions has allowed us to put also a strong upper bound on $\beta$, 
which comes from the requirement of validity of a perturbative treatment. 
This severely restricts the space of parameters allowed for models with sharp features and suggests that the
exact limit of an infinitely sharp step is inconsistent. Moreover, this bound can be used to compare the ratio of the signal-to-noise ratio
for the three-point function to the one of the two-point function. Our result is that, within the range of validity of the effective approach, 
the two-point function has the highest signal-to-noise ratio, unless the height of the step is extremely small. However, as the
amplitude of the bispectrum is proportional to the height of the step itself, we expect at the same time a smaller value of non-Gaussianity.
This suggests us that if a future experiment will show a statistically significant detection of feature effects in the bispectrum
without an even more significant detection in the power-spectrum, the result would be difficult to explain only in the frame of models with features
in the inflaton potential. The situation would change 
if we consider the contemporary presence of a feature together with a speed of sound $c_s < 1$,
as contributions from the folded configuration can potentially make the bispectrum the leading observable for feature models.

\section*{Acknowledgements}
The authors would like to thank Matteo Fasiello for interesting and useful discussions. 
The work of N.B. and S.M. was partially supported by the ASI/INAF Agreement I/072/09/0 for the Planck LFI Activity of Phase E2. 
We also acknowledge support by the PRIN 2009 project "La Ricerca di non-Gussianit\`a Primordiale". 

\appendix

\section{Damping functions}\label{appA}

\begin{figure}
 \begin{minipage}{0.49\textwidth}
  \includegraphics[scale=0.215]{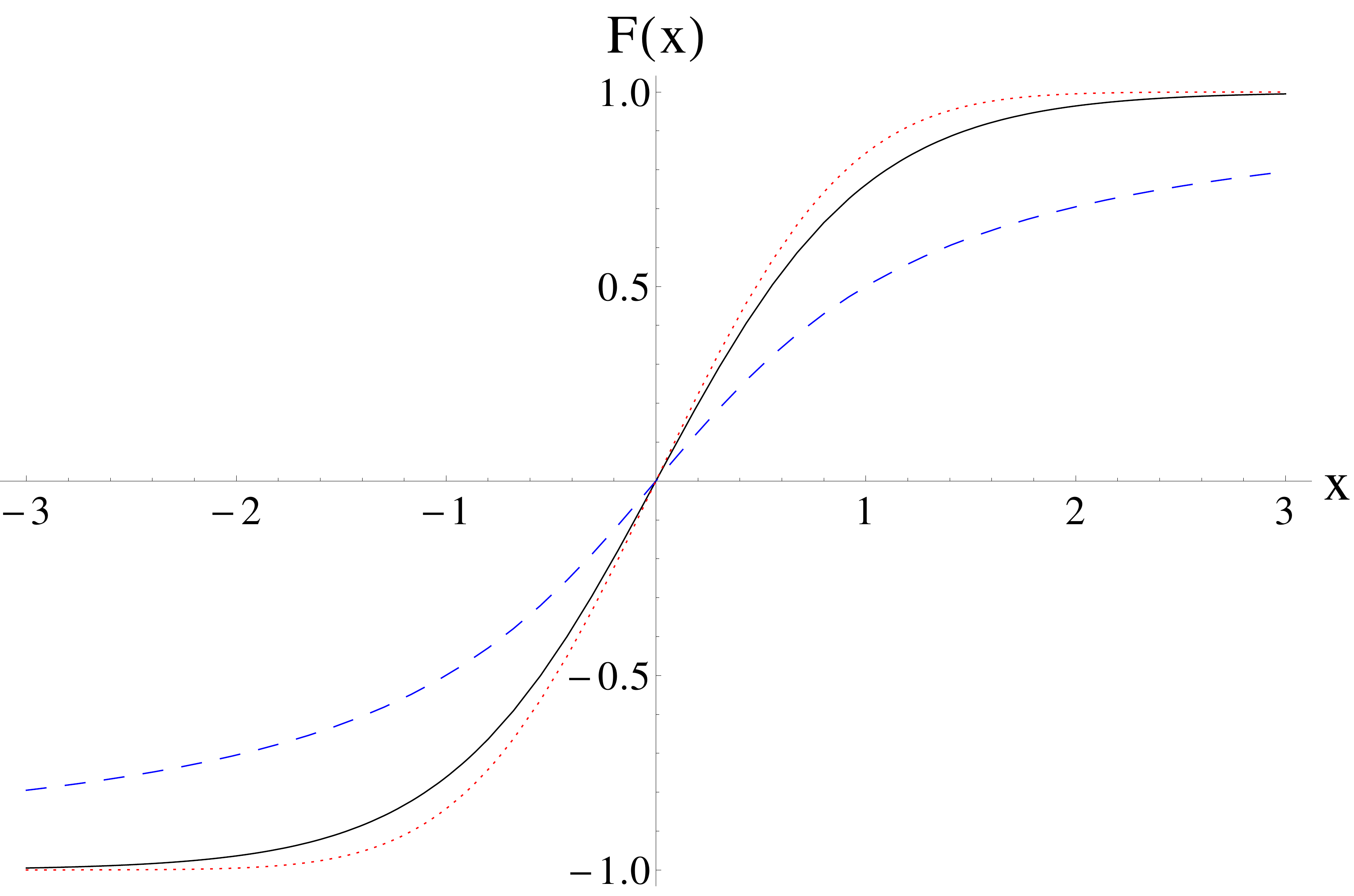}
 \end{minipage}
\begin{minipage}{0.49\textwidth}
 \includegraphics[scale=0.215]{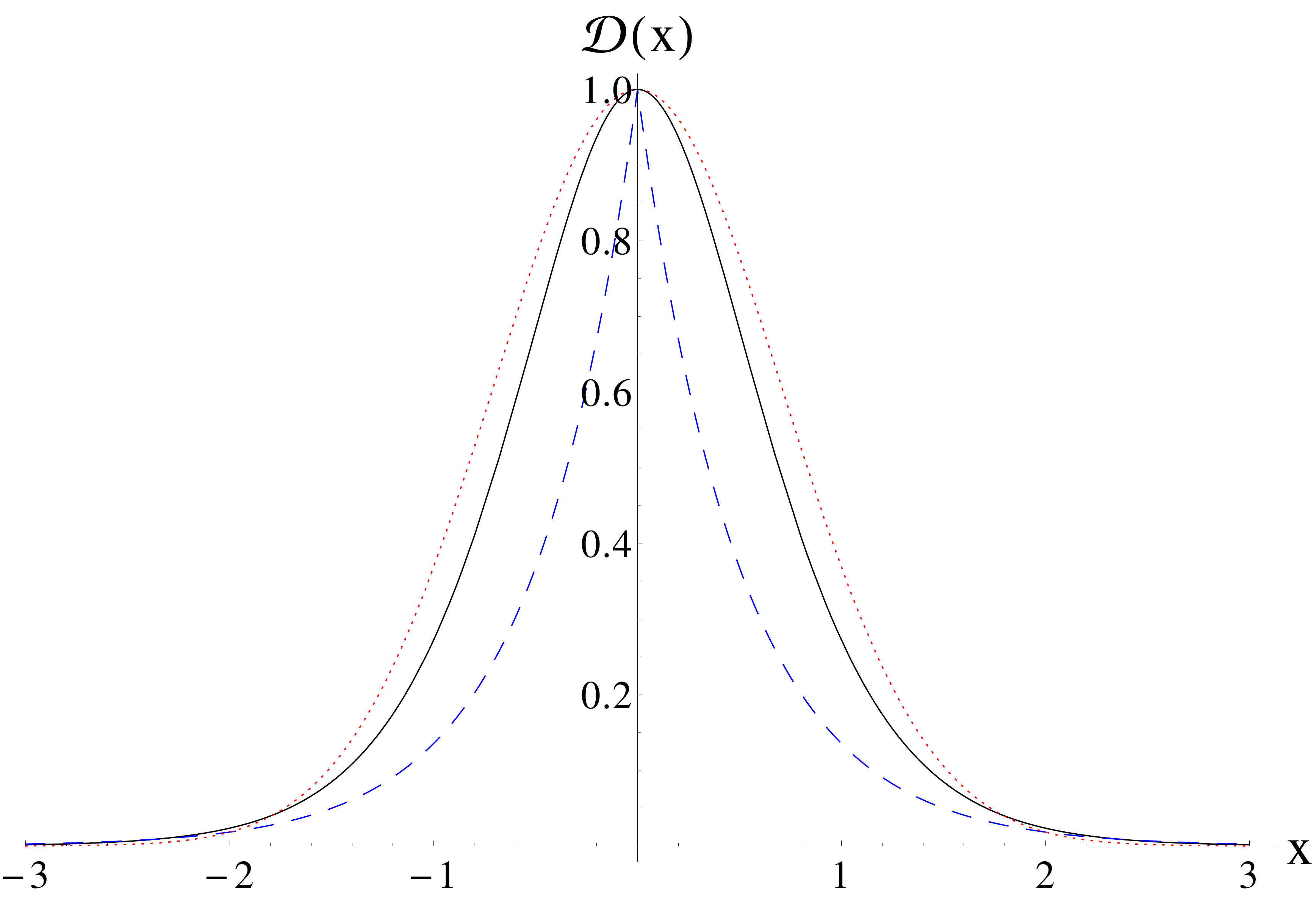}
\end{minipage}
\caption{\footnotesize{Step functions $F(x)$ and damping functions $\mathcal{D}(x)$ for different choice of the step shape, namely
	  hyperbolic tangent \eqref{damp_tanh} (black), arctangent \eqref{damp_arctan} (dashed blue), gaussian integral \eqref{damp_gauss}
	  (dotted red) profiles.}}\label{dampings}
\end{figure}

In this appendix, inspired by \cite{feat:Adshead2011, feat:Stewart2001}
we provide a general treatment for the computation of the typical integrals that can be found when studying models
with features. These integrals in conformal time are generally of the form:
\begin{equation}\label{damp_integral}
 I=\beta \int_{-\infty}^{+\infty}\dif\ln\tau\ \,p(k\tau)\cos(2 k\tau)F'\left(-\beta\ln(\tau/\tau_f)\right) \; ,
\end{equation}
or with sine instead of cosine and where $p(k\tau)$ is a sum of polynomials.
We use the same notation of eq. \eqref{delta_feat}, where $F'(x)$ denotes the derivative of the step function $F(x)$ with respect to its 
argument and $\beta=1/bH$ is taken to be $\beta\gg1$ in the case of sharp step features.
Notice that, as $\beta\gg1$, the derivative of the step $F'(x)$ is strongly peaked in its central value,
namely $\tau=\tau_f$. The polynomials varies slowly in the small region where $F'(x)$ is non-zero and we can replace them by their value
when $F'(x)$ is peaked, namely $(k\tau)^n\to(k\tau_f)^n$. Then, we can use the exponential form of
sine and cosine and change variable to $y=-\beta\ln(\tau/\tau_f)$ to obtain
\begin{equation}
 \frac{1}{2}p(k\tau_f)\left[\int_{-\infty}^{+\infty}\dif y\,\exp\left\{2\mi k \tau_f\me^{-y/\beta}\right\}F'(y)+
 \int_{-\infty}^{+\infty}\dif y\,\exp\left\{-2\mi k \tau_f\me^{-y/\beta}\right\}F'(y)\right] \; .
\end{equation}
Now we linearize the exponential, $\exp(-y/\beta)\simeq1-y/\beta$, to give:
\begin{equation}
 \frac{1}{2}p(k\tau_f)\left[\me^{2\mi k \tau_f}\int_{-\infty}^{+\infty}\dif y\,\me^{-\frac{2\mi k \tau_f}{\beta}y}F'(y)+
 \me^{-2\mi k \tau_f}\int_{-\infty}^{+\infty}\dif y\,\me^{\frac{2\mi k \tau_f}{\beta}y}F'(y)\right] \; .
\end{equation}
We can make this substitution as long as $y\ll\beta$, that is to say that the validity of the approximation breaks down
for $\tau\ll\tau_f$ . However, since $\tau_f$ is the position of the step in conformal time, this corresponds to early times or much
before the step, where we expect that the integral is already negligible. Notice now that the two integrals in the previous equation are
actually the same integral: $F'(x)$ is even , being the derivative of the step $F(x)$, which is an odd function. As a consequence,
we can reconstruct the cosine in front of the integral and write
\begin{equation}
 I=p(k\tau_f)\cos(2k\tau_f)\int_{-\infty}^{+\infty}\dif y\,\me^{-\frac{2\mi k \tau_f}{\beta}y}F'(y) \; .
\end{equation}
It is easy to recognize the Fourier transform of the derivative of the step with respect to the variables $y$ and $2k\tau_f/\beta$,
which is nothing else that the Fourier transform of the step itself
\begin{equation}
 I=2\,p(k\tau_f)\cos(2k\tau_f)\left(\frac{2\mi k \tau_f}{\beta}\hat{\mathcal{F}}\big[F(y)\big]\right)=
 2\,p(k\tau_f)\cos(2k\tau_f)\mathcal{D}\left(\frac{k\tau_f}{\beta}\right) \; .
\end{equation}
We have obtained an oscillating function (with sine or cosine), times a damping envelope $\mathcal{D}$ which is normalized to one.
The further factor $2$ is due to the fact that $F(x)$ goes from $-1$ to $+1$.
This is a quite general result that depends only on the assumption of a very small and very sharp step.
It is also reminiscent of the classical quantum mechanics problem of a potential barrier, where the reflection probability is proportional
to the Fourier transform of the barrier itself.
For different choice of the step shape, we obtain different damping effects (see \imgref{dampings}):
\begin{eqnarray}
 F(x)=\tanh(x) & \qquad \Longrightarrow \qquad & \mathcal{D}\left(\frac{k\tau_f}{\beta}\right)=
    \frac{\pi k \tau_f/\beta}{\sinh\left(\pi k\tau_f/\beta\right)} \; , \label{damp_tanh}\\
 F(x)=\frac{2}{\pi}\arctan(x) & \qquad \Longrightarrow \qquad & \mathcal{D}\left(\frac{k\tau_f}{\beta}\right)=
    \me^{-2\left|\frac{k\tau_f}{\beta}\right|} \; , \label{damp_arctan}\\
 F(x)=\frac{2}{\sqrt{\pi}}\int_{-\infty}^{+\infty}\me^{-x^2} & \qquad \Longrightarrow \qquad & \mathcal{D}\left(\frac{k\tau_f}{\beta}\right)=
    \me^{-\left(\frac{k\tau_f}{\beta}\right)^2} \; \label{damp_gauss}.
\end{eqnarray}
Finally, consider the case $\beta\to\infty$, which is the case of an infinitely sharp step. This corresponds to a step function
in the form of an Heaviside function, whose derivative is a Dirac delta function. In this case the integral \eqref{damp_integral} is
straightforward and correspond to take $\tau=\tau_f$ everywhere. We can see explicitly that no damping envelope arises and oscillations
persist in all $k$-space.

\section{Comparison with previous results and higher-order slow-roll corrections}\label{appB}

In this appendix we show a comparison of our bispectrum \eqref{bs} with that obtained in the previous literature and in particular in the 
GSR approach \cite{feat:Adshead2011}. First of all, consider the third-order action \eqref{S3}. With an integration by parts, it can be
rewritten as
\begin{equation}
 S_3=\int\dif^4x a^3 M_{Pl}^2 \left( \frac{1}{2}\ddot{H}\frac{\delta \mathcal{L}}{\delta \pi}\bigg|_1
    + \frac{1}{2}\dddot{H}\pi^2\dot{\pi}\right)
    -\int\dif^4x M_{Pl}^2\frac{\dif}{\dif t}\left(a^3 \ddot{H}\pi^2\dot{\pi}\right) \; ,
\end{equation}
where we have grouped the terms into the linear equation of motion and a boundary term. It is well known \cite{ng:Arroja2011,
ng:Burrage2011, feat:Adshead2011b} that a field redefinition takes care of all these terms, leaving us with the task of computing
the three-point function only for the operator $\pi^2\dot{\pi}$. Notice now that
\begin{equation}
 \dddot{H}=-\epsilon H^4 \frac{\dif \delta}{\dif\ln\tau} \; ,
\end{equation}
where we have used \eqref{H_feat} and \eqref{delta_feat}. Using the in-in formalism, we find the same expression of eq. \eqref{inin},
where however, the integral over conformal time is to be replaced by the new integral:
\begin{equation}
 \int_{-\infty}^0 \dif\ln\tau \frac{1}{3\tau^3}\,\frac{\dif \delta}{\dif\ln\tau}\left(\frac{\dif}{\dif\ln\tau}\pi^3\right) \; .
\end{equation}
Using \eqref{mode_classic} for the modes and taking permutation and conjugates, we end up with a bispectrum in the form:
\begin{equation}\label{bs_gsr}
 \frac{\mathcal{G}(k_1,k_2,k_3)}{k_1 k_2 k_3}=\frac{1}{8 k_1 k_2 k_3} \left[ -I_0(K)k_1 k_2 k_3 -I_1(K)\sum_{i\neq j}k_i^2k_j+
    I_2(K)K(k_1^2+k_2^2+k_3^2)\right] \; ,
\end{equation}
where $K=k_1+k_2+k_3$ and
\begin{equation}
 I_n=\int\dif\ln\tau\,\frac{\dif \delta}{\dif\ln\tau}W_n(K\tau) \; ,
\end{equation}
\begin{equation}
 W_0(x)=x \sin x\;, \qquad W_1(x)=\cos x\;, \qquad W_2(x)=\frac{\sin x}{x} \; .
\end{equation}
This is formally identical to the GSR form of the bispectrum \cite{feat:Adshead2011, feat:Adshead2011b, feat:Adshead2013}.
One can solve it in the same way as it has been done
for the power spectrum \eqref{ps_integral}, performing an integration by parts and then using the results of appendix \ref{appA}.
\begin{figure}
 \begin{minipage}{0.49\textwidth}
  \includegraphics[scale=0.213]{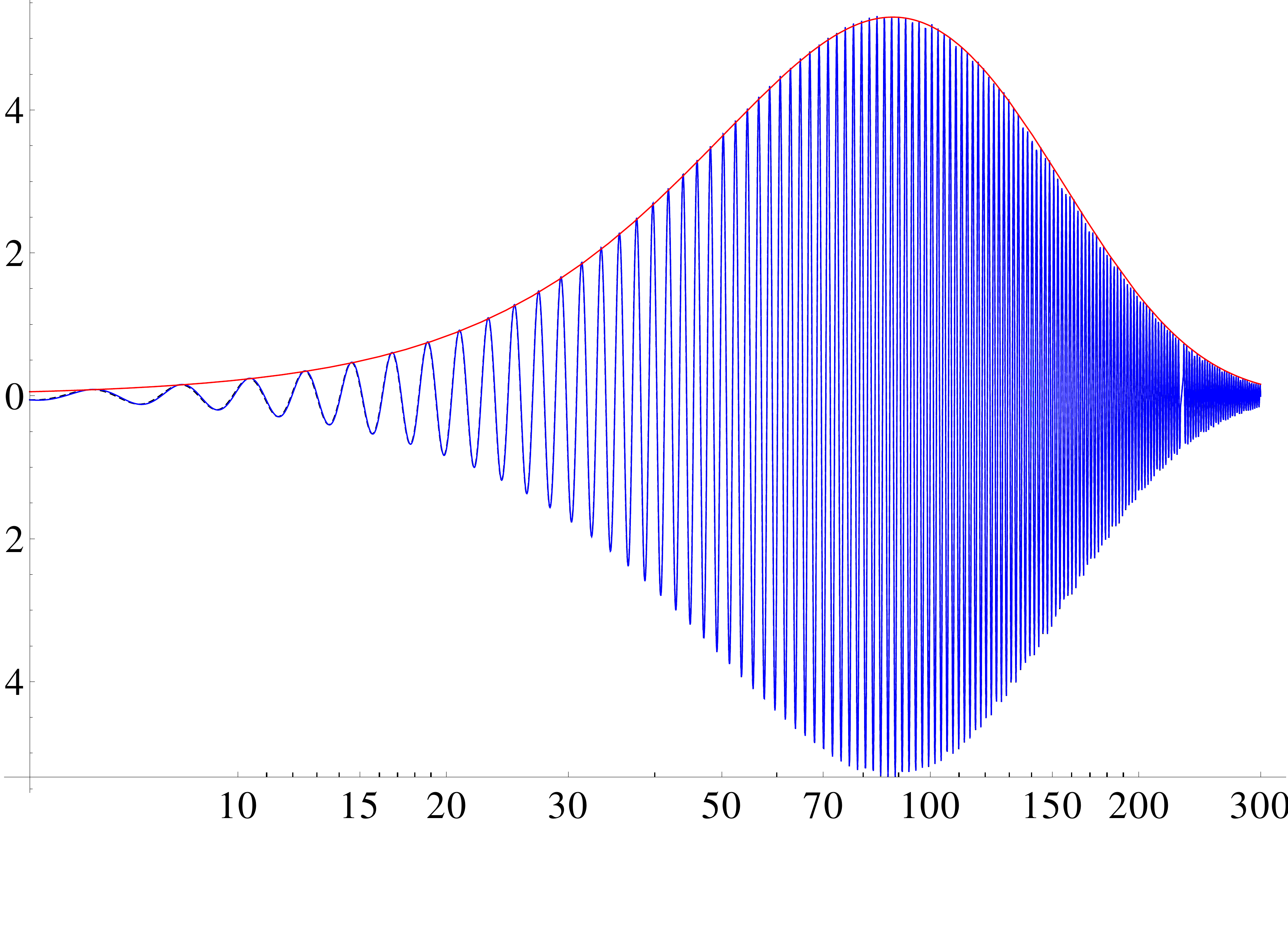}
 \end{minipage}
\begin{minipage}{0.49\textwidth}
  \includegraphics[scale=0.215]{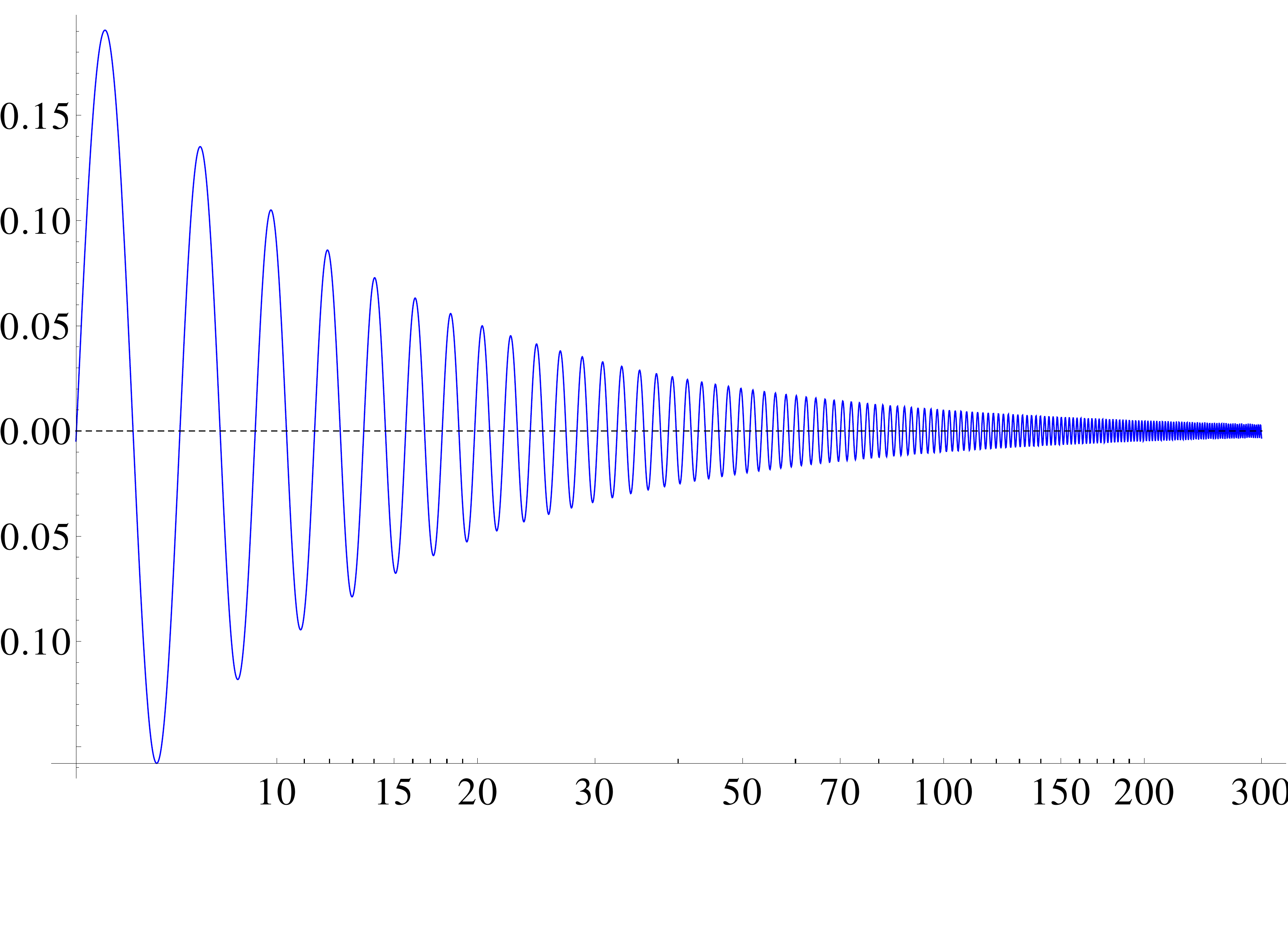}
\end{minipage}
\caption{\footnotesize{In the left panel, plot of  $\frac{\mathcal{G}(k_1,k_2,k_3)}{k_1k_2k_3}$ \eqref{bs_equil} (blue) and the GSR
	  bispectrum \cite{feat:Adshead2011} (dashed red) in the equilateral limit as function of $-k \tau_f$ in the hyperbolic tangent
	  case \eqref{damp_tanh}, together with the envelope profile. In the right panel, difference between the two, divided by the envelope.
	  The choice for the parameters are $\beta=43\pi$ and $\epsilon_{step}=0.001$ for illustration purposes.}}\label{comparison_equil}
\end{figure}
\begin{figure}
 \begin{minipage}{0.49\textwidth}
  \vspace{0.32cm}\includegraphics[scale=0.215]{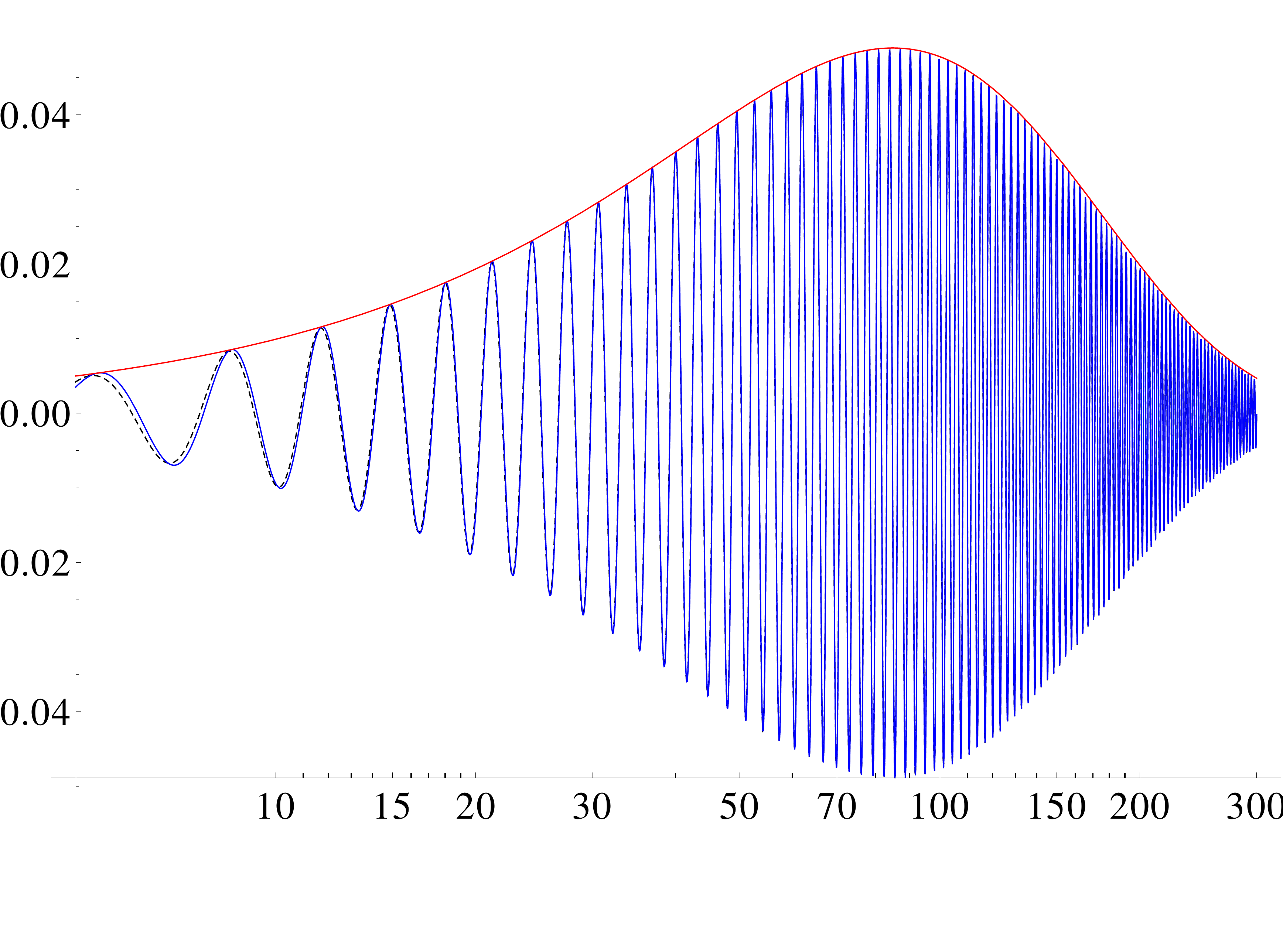}
 \end{minipage}
\begin{minipage}{0.49\textwidth}
  \includegraphics[scale=0.2148]{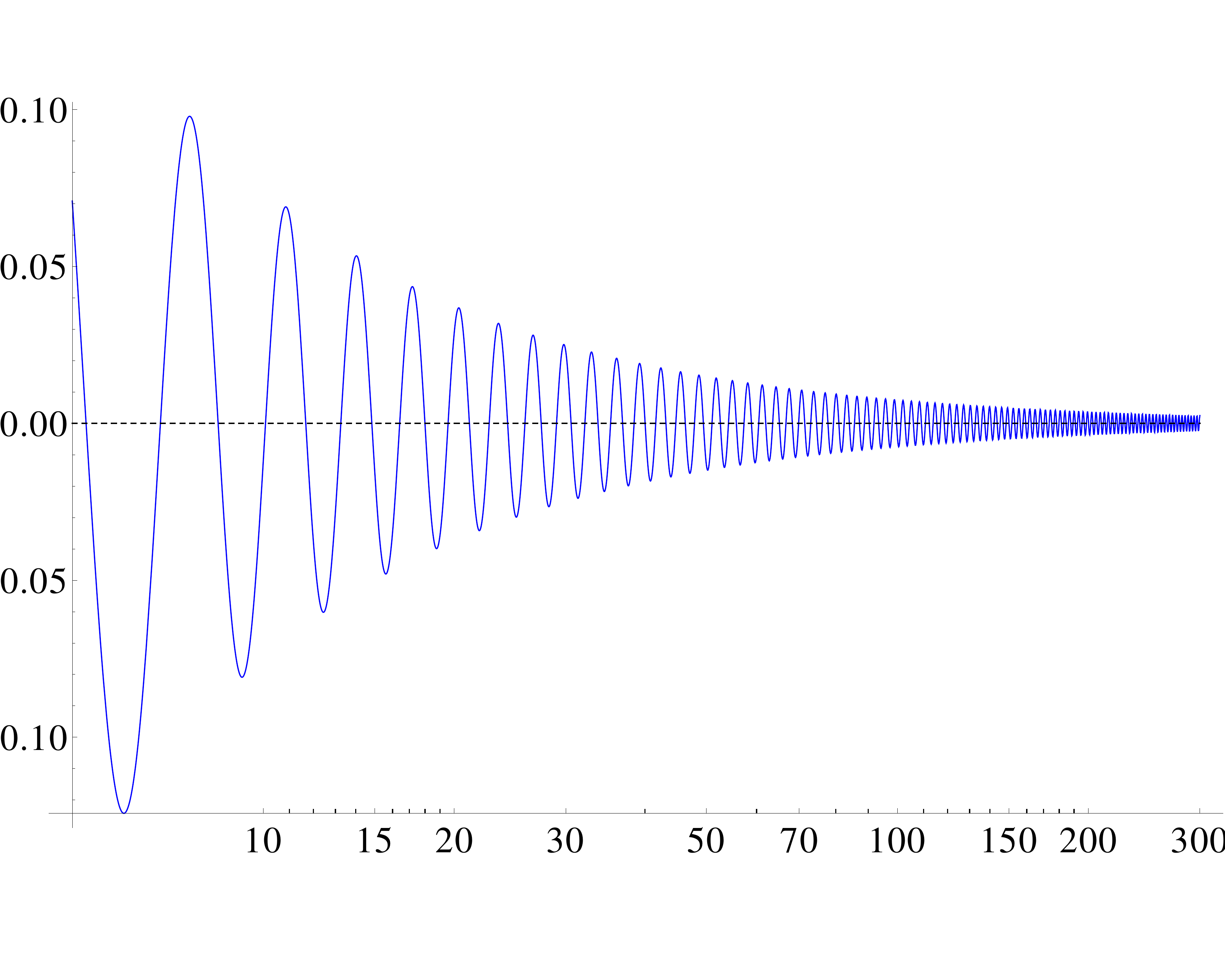}
\end{minipage}
\caption{\footnotesize{In the left panel, plot of  $\frac{\mathcal{G}(k_1,k_2,k_3)}{k_1k_2k_3}$ \eqref{bs3d} in the squeezed limit (blue)
	  and the GSR bispectrum \cite{feat:Adshead2011} (dashed red) in the squeezed limit as function of $-k \tau_f$
	  in the hyperbolic tangent case \eqref{damp_tanh}, together with the envelope profile. In the right panel, difference between the two, divided by the
	  envelope. The choice for the parameters are $\beta=43\pi$ and $\epsilon_{step}=0.001$ for illustration purposes.}}
	  \label{comparison_squeezed}
\end{figure}
As it can be seen from \imgref{comparison_equil} and \imgref{comparison_squeezed}, the bispectra that we have computed are equal to the
GSR ones to very good approximation. The only difference, of the order of $10-15\%$, arises for very large scales, $k\to0$, where however
the non-Gaussianity will be too small to be appreciable. We conclude that our analysis correctly reproduces the effects of sharp step
features.

Also the bispectrum for models with features in the speed of sound can be written in a form similar to eq. \eqref{bs_gsr}.
As we saw, the bispectrum \eqref{bs_cs} has been obtained from eq. \eqref{inin} with the interaction Hamiltonian given by the operator
in eq. \eqref{operator_cs}, with the integration procedure described in appendix \ref{appA}. Let us do here an intermediate step,
considering the integral
\begin{equation}
 I(k_1,k_2,k_3,\tau)=\int_{-\infty}^0\frac{\dif s}{s^2}\sigma\left(-\beta_s\ln(s/s_f)\right)
    \left[\pi_{k_1}(s)^*\pi'_{k_2}(s)^*\pi'_{k_3}(s)^*+\mbox{perm.}+\mbox{c.c.}\right] \; ,
\end{equation}
which is nothing else that the commutator of the in-in formalism after extracting all the factors that will reconstruct the square of
the power spectrum in the final bispectrum \eqref{bs_cs}. Using the expression of the mode functions \eqref{mode_cs}, we obtain
\begin{eqnarray}
 I(k_1,k_2,k_3,\tau_f) = \frac{1}{k_1^3k_2^3k_3^3} & \displaystyle\left[k_1k_2k_3\sum_{i\neq j}k_ik_j\int_{-\infty}^{+\infty}\dif\ln s\,\sigma\left(-\beta_s\ln(s/s_f)\right)s^2 \cos(K s)+ \right.& \nonumber \\
			& \displaystyle \left.+\sum_{i\neq j}k_i^2k_j\int_{-\infty}^{+\infty}\dif\ln s\,\sigma\left(-\beta_s\ln(s/s_f)\right)s\, \sin(K s)\right] & \nonumber \\ 
\end{eqnarray}
Then one can integrate by parts the two pieces of $I(k_1,k_2,k_3,\tau_f)$, sum them and arrive to the expression
\begin{equation}\label{bs_cs_gsr}
 \frac{\mathcal{G}(k_1,k_2,k_3)}{k_1 k_2 k_3}=\frac{1}{8 k_1 k_2 k_3} \left[ T_0(k_1,k_2,k_3)J_0(K) +T_1(k_1,k_2,k_3)J_1(K)\right] \; ,
\end{equation}
where
\begin{equation}
 J_n=\int\dif\ln s\,\frac{\dif \sigma}{\dif\ln s}Z_n(K s) \; ,
\end{equation}
\begin{equation}
 Z_0(x)=\cos x\;, \qquad Z_1(x)=x \sin x \; .
\end{equation}
and
\begin{equation}
 T_0(k_1,k_2,k_3)=\frac{\sum_{i\neq j}k_i^3k_j^2}{{\left(\sum_ik_i\right)}^2}\;,
 \qquad T_1(k_1,k_2,k_3)= \frac{k_1k_2k_3\sum_{i\neq j}k_ik_j}{{\left(\sum_ik_i\right)}^2}\; .
\end{equation}

Notice finally that, given that the forms of the bispectra \eqref{bs}, \eqref{bs_cs} are the same as in the GSR approximation
\eqref{bs_gsr}, as we have shown, then in principle we can use in the same way all the slow-roll corrections computed, for instance, in
\cite{feat:Adshead2011}. As an example, one of these corrections amounts to the addition of the term:
\begin{equation}
 \frac{\mathcal{G}_{corr}(k_1,k_2,k_3)}{k_1k_2k_3}\simeq-\frac{\pi}{16}g_0 I_3(K) \; ,
\end{equation}
\begin{equation}
 I_3(K)=\int_{-\infty}^{+\infty}\dif\ln\tau\,\frac{\dif \delta}{\dif\ln\tau} \,W_3(K\tau_f)\;,\qquad W_3(x)=x \cos x \;,
\end{equation}
where the quantity $g_0$ is a combination of the slow-roll parameters $\epsilon$ \eqref{epsilon_feat}, $\delta$ \eqref{delta_feat} and
$\dif \delta/\dif\ln\tau$ evaluated at $-k\tau=1$:
\begin{equation}
 g_0\simeq6\epsilon-3\delta+\frac{\dif\delta}{\dif\ln\tau} + \mathcal{O}(\epsilon_{step}^2)
\end{equation}
A complete treatment of the possible slow-roll corrections is beyond the scope of this work.

\bibliography{bibliography} \addcontentsline{toc}{section}{References}

\providecommand{\href}[2]{#2}\begingroup\raggedright\begin{thebibliography}{10}

\bibitem{planck_overview}
{\bf Planck} Collaboration, P.~Ade et~al., {\it {Planck 2013 results. I.
  Overview of products and scientific results}},
  \href{http://xxx.lanl.gov/abs/1303.5062}{{\tt arXiv:1303.5062}}.

\bibitem{eft:Cheung2007a}
C.~Cheung, P.~Creminelli, A.~L. Fitzpatrick, J.~Kaplan, and L.~Senatore, {\it
  {The Effective Field Theory of Inflation}},  {\em JHEP} {\bf 0803} (2008)
  014, [\href{http://xxx.lanl.gov/abs/0709.0293}{{\tt arXiv:0709.0293}}].

\bibitem{eft:Creminelli2006}
P.~Creminelli, M.~A. Luty, A.~Nicolis, and L.~Senatore, {\it {Starting the
  Universe: Stable Violation of the Null Energy Condition and Non-standard
  Cosmologies}},  {\em JHEP} {\bf 0612} (2006) 080,
  [\href{http://xxx.lanl.gov/abs/hep-th/0606090}{{\tt hep-th/0606090}}].

\bibitem{eft:Weinberg2008}
S.~Weinberg, {\it {Effective Field Theory for Inflation}},  {\em Phys.Rev.}
  {\bf D77} (2008) 123541, [\href{http://xxx.lanl.gov/abs/0804.4291}{{\tt
  arXiv:0804.4291}}].

\bibitem{feat:Starobinsky1992}
A.~A. Starobinsky, {\it {Spectrum of adiabatic perturbations in the universe
  when there are singularities in the inflation potential}},  {\em JETP Lett.}
  {\bf 55} (1992) 489--494.

\bibitem{feat:Lesgourgues1997}
J.~Lesgourgues, D.~Polarski, and A.~A. Starobinsky, {\it {CDM models with a BSI
  step - like primordial spectrum and a cosmological constant}},  {\em
  Mon.Not.Roy.Astron.Soc.} {\bf 297} (1998) 769--776,
  [\href{http://xxx.lanl.gov/abs/astro-ph/9711139}{{\tt astro-ph/9711139}}].

\bibitem{feat:Wang1999}
L.-M. Wang and M.~Kamionkowski, {\it {The Cosmic microwave background
  bispectrum and inflation}},  {\em Phys.Rev.} {\bf D61} (2000) 063504,
  [\href{http://xxx.lanl.gov/abs/astro-ph/9907431}{{\tt astro-ph/9907431}}].

\bibitem{feat:Adams2001}
J.~A. Adams, B.~Cresswell, and R.~Easther, {\it {Inflationary perturbations
  from a potential with a step}},  {\em Phys.Rev.} {\bf D64} (2001) 123514,
  [\href{http://xxx.lanl.gov/abs/astro-ph/0102236}{{\tt astro-ph/0102236}}].

\bibitem{feat:Gong2005}
J.-O. Gong, {\it {Breaking scale invariance from a singular inflaton
  potential}},  {\em JCAP} {\bf 0507} (2005) 015,
  [\href{http://xxx.lanl.gov/abs/astro-ph/0504383}{{\tt astro-ph/0504383}}].

\bibitem{feat:Peiris2003}
{\bf WMAP} Collaboration, H.~Peiris et~al., {\it {First year Wilkinson
  Microwave Anisotropy Probe (WMAP) observations: Implications for inflation}},
   {\em Astrophys.J.Suppl.} {\bf 148} (2003) 213,
  [\href{http://xxx.lanl.gov/abs/astro-ph/0302225}{{\tt astro-ph/0302225}}].

\bibitem{feat:Hunt2004}
P.~Hunt and S.~Sarkar, {\it {Multiple inflation and the WMAP 'glitches'}},
  {\em Phys.Rev.} {\bf D70} (2004) 103518,
  [\href{http://xxx.lanl.gov/abs/astro-ph/0408138}{{\tt astro-ph/0408138}}].

\bibitem{feat:Covi2006}
L.~Covi, J.~Hamann, A.~Melchiorri, A.~Slosar, and I.~Sorbera, {\it {Inflation
  and WMAP three year data: Features have a Future!}},  {\em Phys.Rev.} {\bf
  D74} (2006) 083509, [\href{http://xxx.lanl.gov/abs/astro-ph/0606452}{{\tt
  astro-ph/0606452}}].

\bibitem{feat:Joy2007}
M.~Joy, V.~Sahni, and A.~A. Starobinsky, {\it {A New Universal Local Feature in
  the Inflationary Perturbation Spectrum}},  {\em Phys.Rev.} {\bf D77} (2008)
  023514, [\href{http://xxx.lanl.gov/abs/0711.1585}{{\tt arXiv:0711.1585}}].

\bibitem{feat:Joy2008}
M.~Joy, A.~Shafieloo, V.~Sahni, and A.~A. Starobinsky, {\it {Is a step in the
  primordial spectral index favored by CMB data ?}},  {\em JCAP} {\bf 0906}
  (2009) 028, [\href{http://xxx.lanl.gov/abs/0807.3334}{{\tt
  arXiv:0807.3334}}].

\bibitem{feat:Mortonson2009}
M.~J. Mortonson, C.~Dvorkin, H.~V. Peiris, and W.~Hu, {\it {CMB polarization
  features from inflation versus reionization}},  {\em Phys.Rev.} {\bf D79}
  (2009) 103519, [\href{http://xxx.lanl.gov/abs/0903.4920}{{\tt
  arXiv:0903.4920}}].

\bibitem{feat:Hazra2010}
D.~K. Hazra, M.~Aich, R.~K. Jain, L.~Sriramkumar, and T.~Souradeep, {\it
  {Primordial features due to a step in the inflaton potential}},  {\em JCAP}
  {\bf 1010} (2010) 008, [\href{http://xxx.lanl.gov/abs/1005.2175}{{\tt
  arXiv:1005.2175}}].

\bibitem{feat:Benetti2011}
M.~Benetti, M.~Lattanzi, E.~Calabrese, and A.~Melchiorri, {\it {Features in the
  primordial spectrum: new constraints from WMAP7+ACT data and prospects for
  Planck}},  {\em Phys.Rev.} {\bf D84} (2011) 063509,
  [\href{http://xxx.lanl.gov/abs/1107.4992}{{\tt arXiv:1107.4992}}].

\bibitem{feat:Hazra2012}
D.~K. Hazra, L.~Sriramkumar, and J.~Martin, {\it {BINGO: A code for the
  efficient computation of the scalar bi-spectrum}},  {\em JCAP} {\bf 1305}
  (2013) 026, [\href{http://xxx.lanl.gov/abs/1201.0926}{{\tt
  arXiv:1201.0926}}].

\bibitem{feat:Adshead2011}
P.~Adshead, C.~Dvorkin, W.~Hu, and E.~A. Lim, {\it {Non-Gaussianity from Step
  Features in the Inflationary Potential}},  {\em Phys.Rev.} {\bf D85} (2012)
  023531, [\href{http://xxx.lanl.gov/abs/1110.3050}{{\tt arXiv:1110.3050}}].

\bibitem{feat:Chen2006}
X.~Chen, R.~Easther, and E.~A. Lim, {\it {Large Non-Gaussianities in Single
  Field Inflation}},  {\em JCAP} {\bf 0706} (2007) 023,
  [\href{http://xxx.lanl.gov/abs/astro-ph/0611645}{{\tt astro-ph/0611645}}].

\bibitem{ng:Chen2008}
X.~Chen, R.~Easther, and E.~A. Lim, {\it {Generation and Characterization of
  Large Non-Gaussianities in Single Field Inflation}},  {\em JCAP} {\bf 0804}
  (2008) 010, [\href{http://xxx.lanl.gov/abs/0801.3295}{{\tt
  arXiv:0801.3295}}].

\bibitem{feat:Barnaby2010a}
N.~Barnaby, {\it {On Features and Nongaussianity from Inflationary Particle
  Production}},  {\em Phys.Rev.} {\bf D82} (2010) 106009,
  [\href{http://xxx.lanl.gov/abs/1006.4615}{{\tt arXiv:1006.4615}}].

\bibitem{feat:Barnaby2010b}
N.~Barnaby, {\it {Nongaussianity from Particle Production During Inflation}},
  {\em Adv.Astron.} {\bf 2010} (2010) 156180,
  [\href{http://xxx.lanl.gov/abs/1010.5507}{{\tt arXiv:1010.5507}}].

\bibitem{feat:Romano2008}
A.~E. Romano and M.~Sasaki, {\it {Effects of particle production during
  inflation}},  {\em Phys.Rev.} {\bf D78} (2008) 103522,
  [\href{http://xxx.lanl.gov/abs/0809.5142}{{\tt arXiv:0809.5142}}].

\bibitem{feat:Ashoorioon2006}
A.~Ashoorioon and A.~Krause, {\it {Power Spectrum and Signatures for Cascade
  Inflation}},  \href{http://xxx.lanl.gov/abs/hep-th/0607001}{{\tt
  hep-th/0607001}}.

\bibitem{feat:Ashoorioon2008}
A.~Ashoorioon, A.~Krause, and K.~Turzynski, {\it {Energy Transfer in Multi
  Field Inflation and Cosmological Perturbations}},  {\em JCAP} {\bf 0902}
  (2009) 014, [\href{http://xxx.lanl.gov/abs/0810.4660}{{\tt
  arXiv:0810.4660}}].

\bibitem{feat:Barnaby2009}
N.~Barnaby and Z.~Huang, {\it {Particle Production During Inflation:
  Observational Constraints and Signatures}},  {\em Phys.Rev.} {\bf D80} (2009)
  126018, [\href{http://xxx.lanl.gov/abs/0909.0751}{{\tt arXiv:0909.0751}}].

\bibitem{feat:Battefeld2010a}
D.~Battefeld, T.~Battefeld, H.~Firouzjahi, and N.~Khosravi, {\it {Brane
  Annihilations during Inflation}},  {\em JCAP} {\bf 1007} (2010) 009,
  [\href{http://xxx.lanl.gov/abs/1004.1417}{{\tt arXiv:1004.1417}}].

\bibitem{feat:Battefeld2010b}
D.~Battefeld, T.~Battefeld, J.~Giblin, John~T., and E.~K. Pease, {\it
  {Observable Signatures of Inflaton Decays}},  {\em JCAP} {\bf 1102} (2011)
  024, [\href{http://xxx.lanl.gov/abs/1012.1372}{{\tt arXiv:1012.1372}}].

\bibitem{feat:Firouzjahi2010}
H.~Firouzjahi and S.~Khoeini-Moghaddam, {\it {Fields Annihilation and Particles
  Creation in DBI inflation}},  {\em JCAP} {\bf 1102} (2011) 012,
  [\href{http://xxx.lanl.gov/abs/1011.4500}{{\tt arXiv:1011.4500}}].

\bibitem{feat:Abolhasani2010}
A.~A. Abolhasani, H.~Firouzjahi, and M.~H. Namjoo, {\it {Curvature
  Perturbations and non-Gaussianities from Waterfall Phase Transition during
  Inflation}},  {\em Class.Quant.Grav.} {\bf 28} (2011) 075009,
  [\href{http://xxx.lanl.gov/abs/1010.6292}{{\tt arXiv:1010.6292}}].

\bibitem{feat:Abolhasani2012}
A.~A. Abolhasani, H.~Firouzjahi, S.~Khosravi, and M.~Sasaki, {\it {Local
  Features with Large Spiky non-Gaussianities during Inflation}},
  \href{http://xxx.lanl.gov/abs/1204.3722}{{\tt arXiv:1204.3722}}.

\bibitem{feat:Saito2013}
R.~Saito and Y.-i. Takamizu, {\it {Localized Features in Non-Gaussianity from
  Heavy Physics}},  {\em JCAP} {\bf 1306} (2013) 031,
  [\href{http://xxx.lanl.gov/abs/1303.3839}{{\tt arXiv:1303.3839}}].

\bibitem{feat:Gao2012}
X.~Gao, D.~Langlois, and S.~Mizuno, {\it {Influence of heavy modes on
  perturbations in multiple field inflation}},  {\em JCAP} {\bf 1210} (2012)
  040, [\href{http://xxx.lanl.gov/abs/1205.5275}{{\tt arXiv:1205.5275}}].

\bibitem{planck_ps}
{\bf Planck} Collaboration, P.~Ade et~al., {\it {Planck 2013 results. XXII.
  Constraints on inflation}},  \href{http://xxx.lanl.gov/abs/1303.5082}{{\tt
  arXiv:1303.5082}}.

\bibitem{planck_bs}
{\bf Planck} Collaboration, P.~Ade et~al., {\it {Planck 2013 Results. XXIV.
  Constraints on primordial non-Gaussianity}},
  \href{http://xxx.lanl.gov/abs/1303.5084}{{\tt arXiv:1303.5084}}.

\bibitem{feat:Chen2008}
R.~Bean, X.~Chen, G.~Hailu, S.-H.~H. Tye, and J.~Xu, {\it {Duality Cascade in
  Brane Inflation}},  {\em JCAP} {\bf 0803} (2008) 026,
  [\href{http://xxx.lanl.gov/abs/0802.0491}{{\tt arXiv:0802.0491}}].

\bibitem{feat:Piazza2008}
J.~Khoury and F.~Piazza, {\it {Rapidly-Varying Speed of Sound, Scale Invariance
  and Non-Gaussian Signatures}},  {\em JCAP} {\bf 0907} (2009) 026,
  [\href{http://xxx.lanl.gov/abs/0811.3633}{{\tt arXiv:0811.3633}}].

\bibitem{feat:Nakashima2010}
M.~Nakashima, R.~Saito, Y.-i. Takamizu, and J.~Yokoyama, {\it {The effect of
  varying sound velocity on primordial curvature perturbations}},  {\em
  Prog.Theor.Phys.} {\bf 125} (2011) 1035--1052,
  [\href{http://xxx.lanl.gov/abs/1009.4394}{{\tt arXiv:1009.4394}}].

\bibitem{feat:Achucarro2010}
A.~Achucarro, J.-O. Gong, S.~Hardeman, G.~A. Palma, and S.~P. Patil, {\it
  {Features of heavy physics in the CMB power spectrum}},  {\em JCAP} {\bf
  1101} (2011) 030, [\href{http://xxx.lanl.gov/abs/1010.3693}{{\tt
  arXiv:1010.3693}}].

\bibitem{feat:Park2012}
M.~Park and L.~Sorbo, {\it {Sudden variations in the speed of sound during
  inflation: features in the power spectrum and bispectrum}},
  \href{http://xxx.lanl.gov/abs/1201.2903}{{\tt arXiv:1201.2903}}.

\bibitem{feat:Achucarro2012a}
A.~Achucarro, J.-O. Gong, S.~Hardeman, G.~A. Palma, and S.~P. Patil, {\it
  {Effective theories of single field inflation when heavy fields matter}},
  {\em JHEP} {\bf 1205} (2012) 066,
  [\href{http://xxx.lanl.gov/abs/1201.6342}{{\tt arXiv:1201.6342}}].

\bibitem{feat:Achucarro2012b}
A.~Achucarro, V.~Atal, S.~Cespedes, J.-O. Gong, G.~A. Palma, et~al., {\it
  {Heavy fields, reduced speeds of sound and decoupling during inflation}},
  {\em Phys.Rev.} {\bf D86} (2012) 121301,
  [\href{http://xxx.lanl.gov/abs/1205.0710}{{\tt arXiv:1205.0710}}].

\bibitem{feat:Miranda2012}
V.~Miranda, W.~Hu, and P.~Adshead, {\it {Warp Features in DBI Inflation}},
  \href{http://xxx.lanl.gov/abs/1207.2186}{{\tt arXiv:1207.2186}}.

\bibitem{feat:Adshead2013}
P.~Adshead, W.~Hu, and V.~Miranda, {\it {Bispectrum in Single-Field Inflation
  Beyond Slow-Roll}},  \href{http://xxx.lanl.gov/abs/1303.7004}{{\tt
  arXiv:1303.7004}}.

\bibitem{ng:Weinberg2005}
S.~Weinberg, {\it {Quantum contributions to cosmological correlations}},  {\em
  Phys.Rev.} {\bf D72} (2005) 043514,
  [\href{http://xxx.lanl.gov/abs/hep-th/0506236}{{\tt hep-th/0506236}}].

\bibitem{res:Behbahani2011}
S.~R. Behbahani, A.~Dymarsky, M.~Mirbabayi, and L.~Senatore, {\it {(Small)
  Resonant non-Gaussianities: Signatures of a Discrete Shift Symmetry in the
  Effective Field Theory of Inflation}},
  \href{http://xxx.lanl.gov/abs/1111.3373}{{\tt arXiv:1111.3373}}.

\bibitem{feat:Arroja2011}
F.~Arroja, A.~E. Romano, and M.~Sasaki, {\it {Large and strong scale dependent
  bispectrum in single field inflation from a sharp feature in the mass}},
  {\em Phys.Rev.} {\bf D84} (2011) 123503,
  [\href{http://xxx.lanl.gov/abs/1106.5384}{{\tt arXiv:1106.5384}}].

\bibitem{feat:Dvorkin2009}
C.~Dvorkin and W.~Hu, {\it {Generalized Slow Roll for Large Power Spectrum
  Features}},  {\em Phys.Rev.} {\bf D81} (2010) 023518,
  [\href{http://xxx.lanl.gov/abs/0910.2237}{{\tt arXiv:0910.2237}}].

\bibitem{feat:Stewart2001}
E.~D. Stewart, {\it {The Spectrum of density perturbations produced during
  inflation to leading order in a general slow roll approximation}},  {\em
  Phys.Rev.} {\bf D65} (2002) 103508,
  [\href{http://xxx.lanl.gov/abs/astro-ph/0110322}{{\tt astro-ph/0110322}}].

\bibitem{ng:Maldacena2002}
J.~M. Maldacena, {\it {Non-Gaussian features of primordial fluctuations in
  single field inflationary models}},  {\em JHEP} {\bf 0305} (2003) 013,
  [\href{http://xxx.lanl.gov/abs/astro-ph/0210603}{{\tt astro-ph/0210603}}].

\bibitem{ng:Creminelli2004}
P.~Creminelli and M.~Zaldarriaga, {\it {Single field consistency relation for
  the 3-point function}},  {\em JCAP} {\bf 0410} (2004) 006,
  [\href{http://xxx.lanl.gov/abs/astro-ph/0407059}{{\tt astro-ph/0407059}}].

\bibitem{eft:Cheung2007b}
C.~Cheung, A.~L. Fitzpatrick, J.~Kaplan, and L.~Senatore, {\it {On the
  consistency relation of the 3-point function in single field inflation}},
  {\em JCAP} {\bf 0802} (2008) 021,
  [\href{http://xxx.lanl.gov/abs/0709.0295}{{\tt arXiv:0709.0295}}].

\bibitem{ng:Creminelli2011}
P.~Creminelli, G.~D'Amico, M.~Musso, and J.~Norena, {\it {The (not so) squeezed
  limit of the primordial 3-point function}},  {\em JCAP} {\bf 1111} (2011)
  038, [\href{http://xxx.lanl.gov/abs/1106.1462}{{\tt arXiv:1106.1462}}].

\bibitem{ng:Chen2013}
X.~Chen, H.~Firouzjahi, M.~H. Namjoo, and M.~Sasaki, {\it {A Single Field
  Inflation Model with Large Local Non-Gaussianity}},  {\em Europhys.Lett.}
  {\bf 102} (2013) 59001, [\href{http://xxx.lanl.gov/abs/1301.5699}{{\tt
  arXiv:1301.5699}}].

\bibitem{ng:Senatore2012}
L.~Senatore and M.~Zaldarriaga, {\it {A Note on the Consistency Condition of
  Primordial Fluctuations}},  {\em JCAP} {\bf 1208} (2012) 001,
  [\href{http://xxx.lanl.gov/abs/1203.6884}{{\tt arXiv:1203.6884}}].

\bibitem{eft:Creminelli2013}
P.~Creminelli, A.~Perko, L.~Senatore, M.~Simonovi\'c, and G.~Trevisan, {\it
  {The Physical Squeezed Limit: Consistency Relations at Order $q^2$}},
  \href{http://xxx.lanl.gov/abs/1307.0503}{{\tt arXiv:1307.0503}}.

\bibitem{eft:Agarwal2012}
N.~Agarwal, R.~Holman, A.~J. Tolley, and J.~Lin, {\it {Effective field theory
  and non-Gaussianity from general inflationary states}},
  \href{http://xxx.lanl.gov/abs/1212.1172}{{\tt arXiv:1212.1172}}.

\bibitem{ng:Flauger2013}
R.~Flauger, D.~Green, and R.~A. Porto, {\it {On Squeezed Limits in Single-Field
  Inflation - Part I}},  \href{http://xxx.lanl.gov/abs/1303.1430}{{\tt
  arXiv:1303.1430}}.

\bibitem{ng:Aravind2013}
A.~Aravind, D.~Lorshbough, and S.~Paban, {\it {Non-Gaussianity from Excited
  Initial Inflationary States}},  {\em JHEP} {\bf 1307} (2013) 076,
  [\href{http://xxx.lanl.gov/abs/1303.1440}{{\tt arXiv:1303.1440}}].

\bibitem{feat:Chen2011a}
X.~Chen, {\it {Primordial Features as Evidence for Inflation}},  {\em JCAP}
  {\bf 1201} (2012) 038, [\href{http://xxx.lanl.gov/abs/1104.1323}{{\tt
  arXiv:1104.1323}}].

\bibitem{ng:Chen2010}
X.~Chen, {\it {Primordial Non-Gaussianities from Inflation Models}},  {\em
  Adv.Astron.} {\bf 2010} (2010) 638979,
  [\href{http://xxx.lanl.gov/abs/1002.1416}{{\tt arXiv:1002.1416}}].

\bibitem{feat:arroja2012}
F.~Arroja and M.~Sasaki, {\it {Strong scale dependent bispectrum in the
  Starobinsky model of inflation}},  {\em JCAP} {\bf 1208} (2012) 012,
  [\href{http://xxx.lanl.gov/abs/1204.6489}{{\tt arXiv:1204.6489}}].

\bibitem{eft:Baumann2011}
D.~Baumann and D.~Green, {\it {Equilateral Non-Gaussianity and New Physics on
  the Horizon}},  {\em JCAP} {\bf 1109} (2011) 014,
  [\href{http://xxx.lanl.gov/abs/1102.5343}{{\tt arXiv:1102.5343}}].

\bibitem{eft:Baumann2011c}
D.~Baumann, L.~Senatore, and M.~Zaldarriaga, {\it {Scale-Invariance and the
  Strong Coupling Problem}},  {\em JCAP} {\bf 1105} (2011) 004,
  [\href{http://xxx.lanl.gov/abs/1101.3320}{{\tt arXiv:1101.3320}}].

\bibitem{eft:Avgoustidis2012}
A.~Avgoustidis, S.~Cremonini, A.-C. Davis, R.~H. Ribeiro, K.~Turzynski, et~al.,
  {\it {Decoupling Survives Inflation: A Critical Look at Effective Field
  Theory Violations During Inflation}},  {\em JCAP} {\bf 1206} (2012) 025,
  [\href{http://xxx.lanl.gov/abs/1203.0016}{{\tt arXiv:1203.0016}}].

\bibitem{eft:Cremonini2010}
S.~Cremonini, Z.~Lalak, and K.~Turzynski, {\it {Strongly Coupled Perturbations
  in Two-Field Inflationary Models}},  {\em JCAP} {\bf 1103} (2011) 016,
  [\href{http://xxx.lanl.gov/abs/1010.3021}{{\tt arXiv:1010.3021}}].

\bibitem{eft:Bartolo2010a}
N.~Bartolo, M.~Fasiello, S.~Matarrese, and A.~Riotto, {\it {Large
  non-Gaussianities in the Effective Field Theory Approach to Single-Field
  Inflation: the Bispectrum}},  {\em JCAP} {\bf 1008} (2010) 008,
  [\href{http://xxx.lanl.gov/abs/1004.0893}{{\tt arXiv:1004.0893}}].

\bibitem{feat:Hu2011}
W.~Hu, {\it {Generalized Slow Roll for Non-Canonical Kinetic Terms}},  {\em
  Phys.Rev.} {\bf D84} (2011) 027303,
  [\href{http://xxx.lanl.gov/abs/1104.4500}{{\tt arXiv:1104.4500}}].

\bibitem{ng:Chen2006}
X.~Chen, M.-x. Huang, S.~Kachru, and G.~Shiu, {\it {Observational signatures
  and non-Gaussianities of general single field inflation}},  {\em JCAP} {\bf
  0701} (2007) 002, [\href{http://xxx.lanl.gov/abs/hep-th/0605045}{{\tt
  hep-th/0605045}}].

\bibitem{res:Chen2010}
X.~Chen, {\it {Folded Resonant Non-Gaussianity in General Single Field
  Inflation}},  {\em JCAP} {\bf 1012} (2010) 003,
  [\href{http://xxx.lanl.gov/abs/1008.2485}{{\tt arXiv:1008.2485}}].

\bibitem{ng:Hu2000}
W.~Hu, {\it {Weak lensing of the CMB: A harmonic approach}},  {\em Phys.Rev.}
  {\bf D62} (2000) 043007,
  [\href{http://xxx.lanl.gov/abs/astro-ph/0001303}{{\tt astro-ph/0001303}}].

\bibitem{ng:Arroja2011}
F.~Arroja and T.~Tanaka, {\it {A note on the role of the boundary terms for the
  non-Gaussianity in general k-inflation}},  {\em JCAP} {\bf 1105} (2011) 005,
  [\href{http://xxx.lanl.gov/abs/1103.1102}{{\tt arXiv:1103.1102}}].

\bibitem{ng:Burrage2011}
C.~Burrage, R.~H. Ribeiro, and D.~Seery, {\it {Large slow-roll corrections to
  the bispectrum of noncanonical inflation}},  {\em JCAP} {\bf 1107} (2011)
  032, [\href{http://xxx.lanl.gov/abs/1103.4126}{{\tt arXiv:1103.4126}}].

\bibitem{feat:Adshead2011b}
P.~Adshead, W.~Hu, C.~Dvorkin, and H.~V. Peiris, {\it {Fast Computation of
  Bispectrum Features with Generalized Slow Roll}},  {\em Phys.Rev.} {\bf D84}
  (2011) 043519, [\href{http://xxx.lanl.gov/abs/1102.3435}{{\tt
  arXiv:1102.3435}}].

\end{thebibliography}\endgroup
\bibliographystyle{JHEP}

\end{document}